\documentclass[12pt,preprint]{aastex}
\usepackage{graphicx}

\def\etal{et\thinspace al.\ }                               

\shortauthors{Cid Fernandes et al.}
\shorttitle{Stellar Populations of LLAGN}

\begin{document}

\title{The Stellar Populations of Low Luminosity Active Galactic
Nuclei. I: Ground Based Observations\footnote{Based on observations
made with the Nordic Optical Telescope, operated on the island of La
Palma jointly by Denmark, Finland, Iceland, Norway, and Sweden, in the
Spanish Observatorio del Roque de los Muchachos of the Instituto de
Astrof\'{\i}sica de Can\'arias.}}

\author{
Roberto Cid Fernandes\altaffilmark{1}, Rosa M. Gonz\'alez Delgado\altaffilmark{2} 
Henrique Schmitt\altaffilmark{3,10,11}, Thaisa Storchi-Bergmann\altaffilmark{4},
Lucimara P. Martins\altaffilmark{5,9}, Enrique P\'erez\altaffilmark{2},
Timothy Heckman\altaffilmark{6,12}, Claus Leitherer\altaffilmark{5}, 
\& Daniel Schaerer\altaffilmark{7,8}}

\affil{(1) \em Depto. de F\'\i sica-CFM, Universidade Federal de Santa 
Catarina, C.P. 476, 88040-900, Florian\'opolis, 
SC, Brazil (cid@astro.ufsc.br)}
\affil{(2) \em Instituto de Astrof\'{\i}sica de Andalucia (CSIC), P.O. Box 3004, 18080 Granada, Spain (rosa@iaa.es; eperez@iaa.es)}
\affil{(3) \em National Radio Astronomy Observatory, PO Box 0, Socorro, NM 87801 (hschmitt@nrao.edu)}
\affil{(4) \em Instituto de F\'\i sica, Universidad Federal do Rio Grande do Sul, C.P. 15001, 91501-970, 
Poto Alegre, RS, Brazil (thaisa@if.ufrgs.br)}
\affil{(5) \em Space Telescope Institute, 3700 San Martin Drive, Baltimore, MD 21218, USA (martins@stsci.edu; leitherer@stsci.edu)}
\affil{(6) \em Department of Physics \& Astronomy, JHU, Baltimore, MD 21218 (heckman@pha.jhu.edu)}
\affil{(7) \em Observatoire de Geneve, 51, Ch. des Maillettes, CH-1290 Sauverny, Switzerland}
\affil{(8) \em Laboratoire d'Astrophysique,UMR 5572, 14 AV. E. Belin, F-31400 Toulouse, France}
\affil{(9) \em Intituto de Astronom\'\i a, Geof\'\i sica e Ciencias Atmosf\'ericas, 05508-900  Sao Paulo, Brazil}

\altaffiltext{10}{Jansky Fellow} 
\altaffiltext{11}{Visiting Astronomer, Kitt Peak National Observatory,
National Optical Astronomy Observatories, which are operated by AURA,
Inc., under a cooperative agreement with the National Science
Foundation.}
\altaffiltext{12}{Also Adjunct Astronomer at STScI} 











\begin{abstract}

We present a spectroscopic study of the stellar populations of Low
Luminosity Active Galactic Nuclei (LLAGN). Our main goal is to
determine whether the stars who live in the innermost (100 pc-scale)
regions of these galaxies are in some way related to the emission line
properties, which would imply a link between the stellar population
and the ionization mechanism.  High signal to noise, ground based
long-slit spectra in the 3500--5500 \AA\ interval were collected for
60 galaxies: 51 LINERs and LINER/HII Transition Objects, 2 Starburst
and 7 non-active galaxies. In this paper, first of a series, we (1)
describe the sample; (2) present the nuclear spectra; (3) characterize
the stellar populations of LLAGN by means of an empirical comparison
with normal galaxies; (4) measure a set of spectral indices, including
several absorption line equivalent widths and colors indicative of
stellar populations; (5) correlate the stellar indices with emission
line ratios which may distinguish between possible excitation sources
for the gas.

Our main findings are: (1) Few LLAGN have a detectable young ($\la
10^7$ yr) starburst component, indicating that very massive stars do
not contribute significantly to the optical continuum. In particular,
no features due to Wolf-Rayet stars were convincingly detected. (2)
High Order Balmer absorption lines of HI (HOBLs), on the other hand,
are detected in $\sim 40\%$ of LLAGN. These features, which are
strongest in $10^8$--$10^9$ yr intermediate age stellar populations,
are accompanied by diluted metal absorption lines and bluer colors
than other objects in the sample.  (3) These intermediate age
populations are very common ($\sim 50\%$) in LLAGN with relatively
weak [OI] emission ([OI]/H$\alpha \le 0.25$), but rare ($\sim 10\%$)
in LLAGN with stronger [OI]. This is intriguing since LLAGN with weak
[OI] have been previously hypothesized to be ``transition objects'' in
which both an AGN and young stars contribute to the emission-line
excitation. Massive stars, if present, are completely outshone by
intermediate age and old stars in the optical. This happens in at
least a couple of objects where independent UV spectroscopy detects
young starbursts not seen in the optical. (4) Objects with
predominantly old stars span the whole range of [OI]/H$\alpha$ values,
but (5) sources with significant young and/or intermediate age
populations are nearly all ($\sim 90\%$) weak [OI] emitters.

These new findings suggest a link between the stellar populations and
the gas ionization mechanism. The strong-[OI] objects are most likely
true LLAGN, with stellar processes being insignificant. However, the
weak-[OI] objects may comprise two populations, one where the
ionization is dominated by stellar processes and another where it is
either governed by an AGN or by a more even mixture of stellar and AGN
processes. Possible stellar sources for the ionization include weak
starbursts, supernova-remnants and evolved post-starburst
populations. These scenarios are examined and constrained by means of
complementary observations and detailed modeling of the stellar
populations in forthcoming communications.

\end{abstract}


\section{Introduction}

\label{sec:Introduction}

Low Ionization Nuclear Emission-line Regions, better known as LINERs,
were first described as a separate class of galaxies by Heckman
(1980). These objects have characteristically strong low-ionization
forbidden lines of [OI], [NII], [SII] and [OII] that distinguish them
from both Seyfert and Starburst nuclei. Because of their small
luminosities compared to Seyferts and quasars, these objects are often
called Low Luminosity Active Galactic Nuclei (LLAGN).  LLAGN are the
most common form of nuclear activity in the local universe. In the
magnitude limited survey by Ho, Filippenko \& Sargent (1995, 1997,
hereafter HFS97), they comprise $\sim 1/3$ of all galaxies.

What powers LLAGN and how do they fit in the global picture of AGN?
Are they all truly ``dwarf Seyferts'' powered by accretion onto
nearly-dormant supermassive black holes (``dead QSOs'') or can some of
them be explained at least partly in terms of stellar process?

These questions have been at the forefront of AGN research for over
two decades. Nowadays there is no doubt that a substantial fraction of
these nuclei contain {\it bona fide} AGN.  Among the most solid
evidence in support of this ``AGN model'' is the fact that $\sim 1/5$
of LLAGN have broad (few thousand km/s) H$\alpha$ emission (HFS97),
signaling the presence of a Broad Line Region, a hallmark of AGN. In
fact, some of the most compelling evidence for the existence of
nuclear accretion disks come from LLAGN with double-peaked broad
Balmer lines (eg, NGC 1097, Storchi-Bergmann \etal 1997; M81, Bower
\etal 1996). X-ray observations (Terashima \etal 2000; Ho \etal 2001)
and the detection of compact nuclear radio emission (Falcke \etal
2000; Nagar \etal 2000) also support the AGN model.  Similarities
between LLAGN and Seyferts were also identified in the properties of
their host galaxies (Ho, Filippenko \& Sargent 2003). In the
theoretical front, it has long been known that photoionization by a
typical AGN can reproduce the characteristic LINER emission line
spectrum (Ferland \& Netzer 1983; Halpern \& Steiner 1983). These and
other evidence (see Barth 2002 and references therein) have
contributed to the general understanding that many, maybe most LLAGN
are {\it bona fide} members of the AGN family, the low luminosity
cousins of Seyferts.

Despite this general consensus, for nearly as long as LINERs have
existed in the literature, it has been suspected that they constitute
a rather mixed bag of phenomena (e.g., Filippenko 1996; Heckman
1996). LINER-like spectra are known to occur in as diverse
environments as cluster cooling flows (e.g.\ Heckman \etal 1989;
Donahue \& Voit 1991), starburst driven super-winds (Heckman, Armus,
\& Miley 1990), and the bulges of early-type galaxies (e.g.\ Heckman
1996) besides the nuclei of galaxies. It is therefore likely that a
variety of ionization mechanisms are capable of producing a LINER-type
spectrum, and indeed several models not involving photoionization by
an AGN have been proposed to explain the emission line characteristics
of these objects.  Among them are shocks (Heckman 1980; Dopita \&
Sutherland 1995), which may be associated with SN remnants,
photoionization by unusually hot young stars (Filippenko \& Terlevich
1992; Shields 1992; Barth \& Shields 2000), or by post-AGB stars
(Binette \etal 1994; Taniguchi, Shioya \& Murayama 2000).  Even in
nuclei where an AGN is conclusively known to exist, it is unclear what
is its role in the gas ionization and in the overall energetics, since
the AGN may coexist with one or more of these other ionizing agents.
For instance, CIV and SiIV PCygni lines from O stars were detected in
the space-UV for a few LLAGN by Maoz \etal (1998), lending support to
the idea that at least some of these objects may be powered by young
stars. Yet, in some of these same nuclei, like NGC 4569 and NGC 4594,
an AGN has been detected by Chandra observations (Ho \etal
2001). These nuclei therefore contain both a compact starburst and an
AGN.  NGC 4303, recently studied by Colina \etal (2002) and
Jim\'enez-Bailon \etal (2003) is another example.  This coexistence of
a starburst and an AGN is reminiscent of the starburst + Seyfert 2
composites discussed by Heckman \etal (1997), Gonz\'alez Delgado \etal
(1998), Storchi-Bergmann, Cid Fernandes \& Schmitt (1998), Gonz\'alez
Delgado, Heckman \& Leitherer (2001), Cid Fernandes \etal (2001a) and
Joguet \etal (2001).

The question then arises of what are the relative roles of accretion
and other processes in the energetics of LLAGN, and how to segregate
nuclei where accretion is the dominant energy source from those
whose emission lines are powered mainly by stars or their remnants.
HFS97 addressed this issue by dividing LLAGN into two subclasses:
``pure'' or ``normal'' LINERs and LINER/HII ``Transition Objects''
(hereafter TOs), whose emission line ratios are intermediate between
those of LINERs and HII regions. The fundamental distinction between
LINERs and TOs is the relative strength of the [OI]$\lambda$6300 line
(e.g.\ [OI]/H$\alpha$). This collisionally-excited line arises in warm
neutral gas, and as such is weak in gas that is photoionized by the
soft radiation from O stars, since the transition between the external
cold neutral gas and internal ionized layers is very sharp. In
contrast, an AGN produces highly energetic photons that penetrate
deeply into (and heat) the neutral gas. Although useful, the division
of LLAGN onto LINERs and TOs is largely arbitrary, given that there is
no clear-cut gap between LINERs and HII nuclei in the [OI]/H$\alpha$
ratio.

The idea when the TO class was introduced was, in the words of HFS97,
that ``TOs can be most naturally explained as normal LINERs whose
integrated spectra are diluted or contaminated by neighboring HII
regions'' (see also Ho, Filippenko \& Sargent 1993; Gon\c{c}alvez,
V\'eron \& V\'eron-Cetty 1999). Accordingly, the most popular models
for these sources involve photoionization by hot stars in a weak
nuclear starburst (Shields 1992; Filippenko \& Terlevich 1992; Barth
\& Shields 2000), although mechanical heating by supernovae following
a short-duration burst (Engelbracht \etal 1998; Alonso Herrero \etal
2000) and photoionization by evolved ``post-starburst'' populations
(Taniguchi \etal 2000) have also been invoked to explain their
emission line properties.

Given that so many different mechanisms are capable of producing a
similar emission line spectrum, it is evident that emission line data
{\it per se} will not be able to distinguish between different models.

In this paper and others in this series we examine the stellar
populations in a sample of LLAGN in search for clues of their nature.
A simple motivation for this is that all alternative non-AGN models
discussed above share a common trait: They all involve stars in one
way or another, be they young and massive (a starburst), older and not
so massive (post-starburst models) or massive but dead (SNR
shocks). Characterizing the stellar populations of LLAGN may thus
provide us with the extra information needed to assess the relevance
of the proposed scenarios.

With this central goal in mind, we have carried out a spectroscopic
survey of LINERs and TOs in the 3500--5500 \AA\ interval. This
spectral range contains several stellar population features of
interest, including the WR bump at 4680 \AA\ (an unambiguous signpost
of the presence of young, $< 10^7$ yr massive stars), high order
Balmer absorption lines (HOBLs) of HI (which signal the presence of
evolved starbursts, from $10^7$ up to $10^9$ yr), the 4000 \AA\ break,
the Balmer jump and several metal lines typical of old populations,
such as CaII K, the G-band and others. A detailed scrutiny of these
features by means of stellar population analysis provides a {\it
direct} test of models proposed for LINERs and TOs.  This technique
can be contrasted to the mostly used {\it indirect} approach of
comparing emission line ratios with predictions from photoionization
or shock models.

Our survey is based on the sample of HFS97, but differs from it in
two important aspects: Wavelength coverage and spatial
resolution. Unlike HFS97 we do not reach the red part of the
spectrum, which includes emission lines which are the basis of
spectral classification. On the other hand, the HFS97 spectra start
at 4230 \AA, whereas our data goes down to 3500 \AA.  This near-UV
edge contains several key stellar population diagnostics (like the
HOBLs) which have not been systematically studied in previous LLAGN
surveys. As shown here, results obtained in this spectral range give
us a new perspective on the nature of these galaxies.

In this paper we concentrate in the presentation of the data.
Companion papers will discuss stellar population synthesis analysis
(Gonz\'alez Delgado \etal 2003, hereafter Paper II), spatial gradients
of spectral properties, morphology, emission line properties and
models.  This paper is organized as follows: In \S\ref{sec:Sample} and
\S\ref{sec:Observations} we describe the sample, observations and
reduction process.  In \S\ref{sec:Spectra} the nuclear spectra are
presented.  A suite of properties has been measured from these
spectra, such as equivalent widths and colors. These are presented in
\S \ref{sec:SpectralProperties}, which also compares our measurements
with previously published results.  This section also presents an
objective definition of the pseudo continuum used to measure stellar
population indices in Bica's system.  In \S\ref{sec:Correlations} we
examine the most important finding of this investigation: The strong
connection between stellar population and emission line properties in
LLAGN. Finally, \S\ref{sec:Conclusions} summarizes our results.

\section{Sample}

\label{sec:Sample}

We selected galaxies from the Ho \etal (1995, 1997) sample for this
study because it is the most complete and homogeneous survey of LLAGN
available and provides a representative population of galaxies in the
local universe.  HFS97 divided LLAGN into two subclasses: ``pure'' or
``normal'' LINERs and LINER/HII ``Transition Objects'' (TOs), whose
emission line ratios are intermediate between those of LINERs and HII
regions. Some Seyfert galaxies also match the usual definition of
LLAGN, $L_{H\alpha} < 10^{40}$ erg$\,$s$^{-1}$, but in this paper
``LLAGN'' is used as a synonimous for LINERs and TOs.  HFS97
distinguish LINERs from TOs by the value of the
[OI]$\lambda6300$/H$\alpha$ emission line ratio: They define LINERs as
nuclei with [OI]/H$\alpha \ge 0.17$, while for TOs $0.08 \le $
[OI]/H$\alpha < 0.17$. Because of its sensitivity to the high energy
part of the ionizing spectrum, the [OI]/H$\alpha$ ratio is a good
indicator of the presence of an AGN.

LINERs and TOs are further distinguished from Seyfert and HII nuclei
on the basis of other line ratios: [OIII]$\lambda5007$/H$\beta < 3$,
[NII]$\lambda6583$/H$\alpha \ge 0.6$ and
[SII]$\lambda\lambda6716,6731$/H$\alpha \ge 0.4$. According to these
criteria, the HFS97 survey contains 94 LINERs and 65 TOs. We have
added three other galaxies to these lists: NGC 772 (classified as
``H/T2:'' by HFS97, ie, a hybrid between an HII nucleus and a TO) and
NGC 2685 (a ``S2/T2:'') were grouped among TOs, while NGC 6951
(classified as a Seyfert 2 by HFS97 and as a Seyfert 2/LINER by Perez
\etal 2000) was grouped among LINERs.  With these additions the HFS
survey contains 162 LLAGN.  This sample will hereafter be called the
``HFS sample''.

We have observed a subset of 28 LINERs and 23 TOs, which corresponds
to $\sim 1/3$ of the HFS sample. For comparison purposes, we also
observed 7 normal galaxies and 2 Starburst nuclei, which gives a total
of 60 galaxies.  Table \ref{tab:sample_properties} lists the galaxies
in our sample, along with several properties extracted from HFS97. As
in HFS97, LINERs are listed as either L2 (if they contain only narrow
lines) or L1.9 (if a broad H$\alpha$ was detected), and similarly for
TOs.  The L2/T2 and T2/L2 mixed-classes are used when a galaxy looks
more like a LINER than like a TO and vice-versa, respectively. These
ambiguous cases arise due to inconsistencies between the
classification implied by different line ratios and measurements
uncertainties. About 13\% of the LINERs and TOs in the HFS sample are
in these ambiguous categories (and many more have uncertain
classifications), which shows that the frontier between LINERs and TOs
is a rather fuzzy one.

\subsection{LINERs, TOs, weak and strong-[OI] nuclei: classification issues}

In this paper we will adopt a slightly different classification
criterion for LLAGN. While HFS97 divide LINERs from TOs at
[OI]/H$\alpha = 0.17$, we prefer to place this dividing line at
0.25. To avoid confusion, we shall refer to sources with [OI]/H$\alpha
\le 0.25$ as ``weak-[OI] nuclei'', while [OI]/H$\alpha > 0.25$ sources
are classed ``strong-[OI]''. Our motivation to introduce these
definitions is entirely empirical. We shall see that the [OI]/H$\alpha
= 0.25$ limit represents much better the combined distributions of
[OI]/H$\alpha$ and stellar population properties in LLAGN. Clearly,
this choice is largely irrelevant insofar as emission lines are
concerned, since weak-[OI] nuclei still have [OI]/H$\alpha$ values
intermediate between strong-[OI] and HII nuclei. In other words,
weak-[OI] nuclei are essentially TOs and strong-[OI] are all LINERs.

Of the 162 LLAGN in the HFS sample, 56 (35\%) have [OI]/H$\alpha >
0.25$ and 106 (65\%) have [OI]/H$\alpha \le 0.25$. The corresponding
fractions for our sample are nearly identical: 17 and 34 of our 51
LLAGN fit our definitions strong and weak-[OI] nuclei respectively.

\begin{deluxetable}{lrrrrrrrr}
\tabletypesize{\tiny}
\tablewidth{0pc}
\tablecaption{Sample Properties}
\tablehead{
  \colhead{Galaxy}   &
  \colhead{Spectral} &
  \colhead{Hubble}   &
  \colhead{T-type}   &
  \colhead{v}        &
  \colhead{dist.}    &
  \colhead{pc/$^{\prime\prime}$} &
  \colhead{log L$_{{\rm H}\alpha}$}     &
  \colhead{[OI]/H$\alpha$}       \\
  \colhead{Name}     &
  \colhead{Class}    &
  \colhead{Type}     &
  \colhead{}         &
  \colhead{km$\,$s$^{-1}$} &
  \colhead{Mpc}      &
  \colhead{pc}       &
  \colhead{erg$\,$s$^{-1}$} &
  \colhead{}         }
\startdata
NGC 266    &  L1.9       &  SB(rs)ab         &   2.0  &  4662  &   62.4  &  302  &   39.30    &    0.280c  \cr
NGC 315    &  L1.9       &  E+:              &  -4.0  &  4935  &   65.8  &  319  &   39.55    &    0.590   \cr
NGC 404    &  L2         &  SA(s)0-:         &  -3.0  &   -46  &    2.4  &   11  &   37.63    &    0.170   \cr
NGC 410    &  T2:        &  E+:              &  -4.0  &  5295  &   70.6  &  342  &   39.43    &    0.097u  \cr
NGC 428    &  L2/T2:     &  SAB(s)m          &   9.0  &  1160  &   14.9  &   72  &   36.98L   &    0.190u  \cr
NGC 521    &  T2/H:      &  SB(r)bc          &   4.0  &  5039  &   67.0  &  324  &   39.16    &    0.086c  \cr
NGC 660    &  T2/H:      &  SB(s)a pec       &   1.0  &   852  &   11.8  &   57  &   38.89    &    0.047   \cr
NGC 718    &  L2         &  SAB(s)a          &   1.0  &  1734  &   21.4  &  103  &   38.45    &    0.210   \cr
NGC 772$^1$   &  H/T2:      &  SA(s)b           &   3.0  &  2458  &   32.6  &  158  &   39.24    &    0.026b  \cr
NGC 841    &  L1.9:      &  (R')SAB(s)ab     &   2.1  &  4539  &   59.5  &  288  &   39.11L   &    0.580b  \cr
NGC 1052   &  L1.9       &  E4               &  -5.0  &  1499  &   17.8  &   86  &   39.45r   &    0.710   \cr
NGC 1161   &  T1.9:      &  SA0              &  -2.0  &  1940  &   25.9  &  125  &   38.70    &    0.140u  \cr
NGC 1169   &  L2         &  SAB(r)b          &   3.0  &  2386  &   33.7  &  163  &   38.67    &    0.320   \cr
NGC 2681   &  L1.9       &  (R')SAB(rs)0/a   &   0.0  &   688  &   13.3  &   64  &   38.83L   &    0.190b  \cr
NGC 2685$^2$  &  S2/T2:     &  (R)SB0+pec       &  -1.0  &   820  &   16.2  &   78  &   38.66    &    0.130b  \cr
NGC 2911   &  L2         &  SA(s)0: pec      &  -2.0  &  3183  &   42.2  &  204  &   39.38r   &    0.310   \cr
NGC 3166   &  L2         &  SAB(rs)0/a       &   0.0  &  1344  &   22.0  &  106  &   39.10    &    0.270   \cr
NGC 3169   &  L2         &  SA(s)a pec       &   1.0  &  1234  &   19.7  &   95  &   39.02    &    0.280   \cr
NGC 3226   &  L1.9       &  E2: pec          &  -5.0  &  1321  &   23.4  &  113  &   38.93    &    0.590   \cr
NGC 3245   &  T2:        &  SA(r)0?          &  -2.0  &  1348  &   22.2  &  107  &   39.59    &    0.086c  \cr
NGC 3627   &  T2/S2      &  SAB(s)b          &   3.0  &   727  &    6.6  &   31  &   38.50    &    0.130   \cr
NGC 3705   &  T2         &  SAB(r)ab         &   2.0  &  1018  &   17.0  &   82  &   38.66    &    0.079b  \cr
NGC 4150   &  T2         &  SA(r)0?          &  -2.0  &    43  &    9.7  &   47  &   38.18    &    0.130   \cr
NGC 4192   &  T2         &  SAB(s)ab         &   2.0  &  -142  &   16.8  &   81  &   38.97    &    0.140   \cr
NGC 4438   &  L1.9       &  SA(s)0/a:        &   0.0  &    64  &   16.8  &   81  &   39.37    &    0.270   \cr
NGC 4569   &  T2         &  SAB(rs)ab        &   2.0  &  -235  &   16.8  &   81  &   40.28r   &    0.062   \cr
NGC 4736   &  L2         &  (R)SA(r)ab       &   2.0  &   307  &    4.3  &   20  &   37.81r   &    0.240   \cr
NGC 4826   &  T2         &  (R)SA(rs)ab      &   2.0  &   411  &    4.1  &   19  &   38.87r   &    0.073   \cr
NGC 5005   &  L1.9       &  SAB(rs)bc        &   4.0  &   948  &   21.3  &  103  &   39.47    &    0.650   \cr
NGC 5055   &  T2         &  SA(rs)bc         &   4.0  &   504  &    7.2  &   34  &   38.62r   &    0.170u  \cr
NGC 5377   &  L2         &  (R)SB(s)a        &   1.0  &  1792  &   31.0  &  150  &   39.18    &    0.250   \cr
NGC 5678   &  T2         &  SAB(rs)b         &   3.0  &  1924  &   35.6  &  172  &   39.19    &    0.079   \cr
NGC 5879   &  T2/L2      &  SA(rs)bc?        &   4.0  &   772  &   16.8  &   81  &   38.32    &    0.160   \cr
NGC 5921   &  T2         &  SB(r)bc          &   4.0  &  1479  &   25.2  &  122  &   39.15r   &    0.110   \cr
NGC 5970   &  L2/T2:     &  SB(r)c           &   5.0  &  1965  &   31.6  &  153  &   38.06    &    0.180c  \cr
NGC 5982   &  L2::       &  E3               &  -5.0  &  2904  &   38.7  &  187  &   38.46c   &    0.490u  \cr
NGC 5985   &  L2         &  SAB(r)b          &   3.0  &  2518  &   39.2  &  190  &   38.94    &    0.300   \cr
NGC 6340   &  L2         &  SA(s)0/a         &   0.0  &  1207  &   22.0  &  106  &   38.50    &    0.430b  \cr
NGC 6384   &  T2         &  SAB(r)bc         &   4.0  &  1667  &   26.6  &  128  &   38.12    &    0.150u  \cr
NGC 6482   &  T2/S2::    &  E:               &  -5.0  &  3921  &   52.3  &  253  &   39.23    &    0.130u  \cr
NGC 6500   &  L2         &  SAab:            &   1.7  &  2999  &   39.7  &  192  &   40.31    &    0.230   \cr
NGC 6501   &  L2::       &  SA0+:            &  -0.5  &  2869  &   39.6  &  191  &   38.01c   &    0.870u  \cr
NGC 6503   &  T2/S2:     &  SA(s)cd          &   6.0  &    42  &    6.1  &   29  &   37.56    &    0.080   \cr
NGC 6702   &  L2::       &  E:               &  -5.0  &  4712  &   62.8  &  304  &   38.61c   &    0.620u  \cr
NGC 6703   &  L2::       &  SA0-             &  -2.5  &  2364  &   35.9  &  174  &   38.51    &    0.360u  \cr
NGC 6951$^3$  &  S2         &  SAB(rs)bc        &   4.0  &  1424  &   24.1  &  116  &   39.07    &    0.230   \cr
NGC 7177   &  T2         &  SAB(r)b          &   3.0  &  1147  &   18.2  &   88  &   38.94    &    0.140   \cr
NGC 7217   &  L2         &  (R)SA(r)ab       &   2.0  &   945  &   16.0  &   77  &   38.86    &    0.250   \cr
NGC 7331   &  T2         &  SA(s)b           &   3.0  &   821  &   14.3  &   69  &   38.49    &    0.097u  \cr
NGC 7626   &  L2::       &  E: pec           &  -5.0  &  3422  &   45.6  &  221  &   38.81    &    0.220u  \cr
NGC 7742   &  T2/L2      &  SA(r)b           &   3.0  &  1653  &   22.2  &  107  &   39.07    &    0.130   \cr
\hline
NGC 3367   &  H          &  SB(rs)c          &   5.0  &  3041  &   43.6  &  211  &   40.98    &    0.031   \cr
NGC 6217   &  H          &  (R)SB(rs)bc      &   4.0  &  1362  &   23.9  &  115  &   40.41    &    0.032   \cr
NGC 205    &  normal     &  dE5 pec          &  -5.0  &  -239  &    0.7  &    3  & $<34.79$r  &   \ldots   \cr
NGC 221    &  normal     &  E2               &  -6.0  &  -205  &    0.7  &    3  & $<36.16$r  &   \ldots   \cr
NGC 224    &  normal     &  SA(s)b           &   3.0  &  -298  &    0.7  &    3  &  \ldots    &   \ldots   \cr
NGC 628    &  normal     &  SA(s)c           &   5.0  &   655  &    9.7  &    3  & $<36.69$r  &   \ldots   \cr
NGC 1023   &  normal     &  SB(rs)0-         &  -3.0  &   632  &   10.5  &   51  & $<37.82$r  &   \ldots   \cr
NGC 2950   &  normal     &  (R)SB(r)0        &  -2.0  &  1337  &   23.3  &  204  & $<38.42$r  &   \ldots   \cr
NGC 6654   &  normal     &  (R')SB(s)0/a     &   0.0  &  1823  &   29.5  &  123  & $<38.11$r  &   \ldots   \cr
\enddata
\label{tab:sample_properties}
\tablecomments{Col.\ (1): Galaxy name; Col.\ (2): Spectral
class. Col.\ (3): Hubble type. Col.\ (4): Numerical Hubble type.
Col.\ (5): Radial velocity.  Col.\ (6): Distance. Col.\ (7): Angular
scale.  Col.\ (8): H$\alpha$ luminosity.  Col.\ (9): [OI]/H$\alpha$
flux ratio. All quantities were extracted from HFS97.  Revised values
of log L$_{{\rm H}\alpha}$, taken from Ho \etal (2003), are indicated
by an ``r'' in column 8. See HFS97 for the meaning of other notes on
columns 8 and 9.}  \tablenotetext{1,2}{Grouped among TOs in this
paper.}  \tablenotetext{3}{Grouped with LINERs in this paper (see
P\'erez \etal 2000).}
\end{deluxetable}

\subsection{Comparison with the HFS sample}

\label{sec:Sample_Comparison}

\begin{figure}
\caption{Distances and morphological T types for the strong-[OI]
(triangles, [OI]/H$\alpha > 0.25$) and weak-[OI] LLAGN (circles,
[OI]/H$\alpha \le 0.25$) in the HFS sample. Filled symbols and filled
areas in the histograms indicate objects in our sample. (All data
extracted from HFS97.)}
\label{fig:f1}
\end{figure}                                           

\begin{figure}
\caption{Equivalent width of the G band versus [OI]/H$\alpha$ for
LLAGN in the HFS sample. Filled symbols mark objects in our
sample. The horizontal line at [OI]/H$\alpha = 0.25$ indicates our
border-line between weak and strong-[OI] nuclei. (All data extracted
from HFS97.)}
\label{fig:OurSample_X_HFSSample_O1Ha_WG}
\end{figure}                                           

Although we have drawn objects exclusively from the HFS sample, it is
important to evaluate whether we have introduced any bias in our
sample selection.  In Figure 1 we
compare the distributions of distances ($d$) and morphological types
($T$) in these two samples. Triangles and circles correspond to strong
and weak-[OI] LLAGN respectively, and filled symbols mark objects
which we have observed.  The visual impression from the bottom left
plot is that our galaxies are well mixed within the HFS sample. This
is confirmed by the histograms projected along both axis of Figure
\ref{fig:f1}, where we see that our galaxies
(filled areas) follow reasonably well the distributions of $d$ and $T$
in the full HFS sample (solid lines). In both samples the distribution
of morphological types of weak-[OI] nuclei is skewed towards somewhat
later type hosts than for strong-[OI] nuclei, but with a substantial
overlap, as discussed by Ho \etal (2003).

In Figure \ref{fig:OurSample_X_HFSSample_O1Ha_WG} we compare the G band
equivalent width and the [OI]/H$\alpha$ ratio, both extracted from
HFS97. Again, one sees that our galaxies are well mixed within the
HFS galaxies. This is confirmed by a statistical analysis. The mean
value and rms dispersion of $W$(G~band) is $4.6 \pm 1.2$ for all LLAGN
in HFS97 and $4.1 \pm 1.3$ in our sample.  No biases were detected in
the [OI]/H$\alpha$ ratio either ($0.27 \pm 0.21$ compared to $0.25 \pm
0.19$ for the HFS and our samples respectively), and similar results
apply to other stellar indices and emission line ratios tabulated by
HFS97.  We thus conclude that our sample is representative of the
full HFS sample in terms of emission line and stellar population
properties.

We note in passing that Figure \ref{fig:OurSample_X_HFSSample_O1Ha_WG}
already gives away an important result of this investigation. Most
objects with a weak G band are also weak [OI] emitters. The ``inverted
L'' shape traced by the galaxies in this plot is only spoiled by a
streak of 3 or 4 LINERs with [OI]/H$\alpha \sim 0.6$ and $W$(G band)
$< 4$ \AA. Apart from these objects, the top-left region of Figure
\ref{fig:OurSample_X_HFSSample_O1Ha_WG} is remarkably empty.  Of the
two objects which intrude most in this ``zone of avoidance'', NGC 5195
and IC 239, the latter has only an upper limit on [OI] and both have
uncertain classifications (HFS97). Similar uncertainties affect
several other points in Figure \ref{fig:OurSample_X_HFSSample_O1Ha_WG},
but these do not affect the global shape of the distribution.  Other
versions of this plot are presented and discussed in \S6.

\section{Observations}

\label{sec:Observations}

\begin{deluxetable}{lrrrr}
\tabletypesize{\tiny}
\tablewidth{0pc}
\tablecaption{Log of observations}
\tablehead{
\colhead{Name} &
\colhead{Telescope} &   
\colhead{Date} &  
\colhead{Exposure (s)} & 
\colhead{airmass}}
\startdata
NGC 266  & NOT  & 8-21-01 & 3x1200  & 1.03   \cr
NGC 315  & NOT  & 8-22-01 & 3x1200  & 1.11   \cr
NGC 404  & KPNO & 7-26-01 & 2x1200  & 1.29   \cr
NGC 410  & NOT  & 8-21-01 & 3x1200  & 1.13   \cr
NGC 428  & NOT  & 8-23-01 & 4x1200  & 1.26   \cr
NGC 521  & NOT  & 8-23-01 & 3x1200  & 1.69   \cr
NGC 660  & NOT  & 8-21-01 & 4x1200  & 1.07   \cr
NGC 718  & NOT  & 8-22-01 & 3x1200  & 1.20   \cr
NGC 772  & NOT  & 8-23-01 & 3x1200  & 1.04   \cr
NGC 841  & NOT  & 8-22-01 & 4x1200  & 1.02   \cr
NGC 1052 & NOT  & 8-22-01 & 2x1200  & 1.26   \cr
NGC 1161 & NOT  & 8-23-01 & 2x1200  & 1.06   \cr
NGC 1169 & NOT  & 11-8-02 & 3x1200  & 1.05   \cr
NGC 2681 & NOT  & 11-8-02 & 3x900   & 1.33   \cr
NGC 2685 & NOT  & 11-8-02 & 3x1200  & 1.03   \cr
NGC 2911 & NOT  & 5-3-03  & 3x1200  & 1.12   \cr
NGC 3166 & NOT  & 5-4-03  & 2x1200  & 1.11   \cr
NGC 3169 & NOT  & 5-3-03  & 3x1200  & 1.14   \cr
NGC 3226 & NOT  & 5-14-01 & 3x1200  & 1.05   \cr
NGC 3245 & NOT  & 5-12-01 & 1800+1200+600  & 1.04   \cr
NGC 3627 & NOT  & 5-13-01 & 3x1200  & 1.09   \cr
NGC 3705 & NOT  & 5-14-01 & 3x1200  & 1.12   \cr
NGC 4150 & NOT  & 5-14-01 & 4x1200  & 1.28   \cr
NGC 4192 & NOT  & 5-12-01 & 4x900   & 1.04   \cr
NGC 4438 & NOT  & 5-15-01 & 4x1200  & 1.25   \cr
NGC 4736 & NOT  & 5-3-03  & 3x900   & 1.04   \cr
NGC 4569 & NOT  & 5-12-01 & 3x1200  & 1.11   \cr
NGC 4826 & NOT  & 5-14-01 & 3x1200  & 1.26   \cr
NGC 5005 & NOT  & 5-3-03  & 3x1200  & 1.09   \cr
NGC 5055 & NOT  & 5-15-01 & 4x1200  & 1.15   \cr
NGC 5377 & NOT  & 5-3-03  & 3x1200  & 1.24   \cr
NGC 5678 & NOT  & 5-13-01 & 4x1200  & 1.18   \cr
NGC 5879 & NOT  & 5-14-01 & 4x1200  & 1.21   \cr
NGC 5921 & NOT  & 5-12-01 & 3x1200  & 1.09   \cr
NGC 5970 & NOT  & 8-23-01 & 3x1200  & 1.19   \cr
NGC 5982 & NOT  & 5-4-03  & 3x1200  & 1.28   \cr
NGC 5985 & NOT  & 8-22-01 & 3x1200  & 1.29   \cr
NGC 6340 & KPNO & 7-26-01 & 3x1200  & 1.34   \cr
NGC 6384 & NOT  & 5-13-01 & 3x1200  & 1.12   \cr
NGC 6482 & NOT  & 5-12-01 & 3x1200  & 1.03   \cr
NGC 6500 & NOT  & 5-12-01 & 3x1200+900& 1.02 \cr
NGC 6501 & NOT  & 8-22-01 & 3x1200  & 1.10   \cr
NGC 6503 & NOT  & 5-13-01 & 3x1200  & 1.33   \cr
NGC 6702 & NOT  & 8-21-01 & 3x1200  & 1.06   \cr
NGC 6703 & KPNO & 7-24-01 & 3x1200  & 1.03   \cr
NGC 6951 & KPNO & 7-24-01 & 2x1200  & 1.21   \cr
NGC 7177 & KPNO & 7-25-01 & 3x1200  & 1.20   \cr
NGC 7217 & KPNO & 7-26-01 & 3x1200  & 1.01   \cr
NGC 7331 & KPNO & 7-24-01 & 3x1200  & 1.03   \cr
NGC 7626 & NOT  & 8-23-01 & 3x1200  & 1.40   \cr
NGC 7742 & KPNO & 7-24-01 & 2x1200  & 1.10   \cr
\hline
NGC 3367 & NOT  & 5-13-01 & 3x1200  & 1.04   \cr
NGC 6217 & NOT  & 8-23-01 & 3x1200  & 1.62   \cr
NGC 205  & NOT  & 11-7-02 & 3x900   & 1.03   \cr
NGC 221  & NOT  & 11-7-02 & 2x1200  & 1.22   \cr
NGC 224  & NOT  & 11-7-02 & 2x900+600& 1.08  \cr
NGC 628  & NOT  & 11-7-02 & 3x1200  & 1.03   \cr
NGC 1023 & NOT  & 11-7-02 & 3x900   & 1.10   \cr
NGC 2950 & NOT  & 11-7-02 & 3x900   & 1.21   \cr
NGC 6654 & NOT  & 5-15-01 & 3x1200  & 1.41   \cr
\enddata
\label{tab:log_obs}
\end{deluxetable}

Observations were carried out in four runs at the 2.5m Nordic Optical
Telescope and one run at the Kitt Peak National Observatory 2.1m.  The
NOT observations used the ALFOSC detector with grism number 6 and a
1$^{\prime\prime}$ wide slit, which gives a dispersion of 1.4 \AA\ per
pixel and covers the wavelength range 3500--5500 \AA.  The slit was
oriented along the paralactic angle. Most observations were made under
sub-arcsecond seeing and photometric conditions.  The KPNO
observations were carried out using the Gold Camera with grating
26\,old and a slit 2$^{\prime\prime}$ wide, which gives a dispersion
of 1.24 \AA\ per pixel, and covers the 3400--5500 \AA\ interval. The
slit was oriented in the N--S direction and the observations were done
as close to the meridian passage as possible to avoid light losses by
differential refraction. The weather during the observations was not
photometric, and the seeing was typically 2--3$^{\prime\prime}$. A log
of the observations is given in Table \ref{tab:log_obs}.

The data were reduced using standard IRAF procedures. The individual
frames were overscanned, bias subtracted and flat field divided. The
calibration also followed standard steps. The wavelength scale was
calibrated using an HeNeAr lamp. The flux calibration was done with
observations of spectrophotometric standard stars observed with a slit
width of 10$^{\prime\prime}$.  The individual frames were combined
using crreject task in IRAF to remove cosmic rays. Sky was subtracted
with the background task fitting a polynomium along the spatial light
profile.

\subsection{Nuclear Extractions}

Most of the spectra have excellent signal out to several arcsec from
the nucleus. In Cid Fernandes \etal (2003b, Paper III) we explore this
spatial information to map the stellar populations in the central
regions of these objects, in analogy with previous work for type 2
Seyferts (Cid Fernandes, Storchi-Bergmann \& Schmitt 1998; Joguet
\etal 2001; Raimann \etal 2003). In this paper we concentrate on the
nuclear spectra. Off-nuclear extractions will only be used as
templates for the nuclear spectra (\S\ref{sec:Spectra}).

Nuclear spectra were extracted from the central 1$^{\prime\prime}
\times $1.1$^{\prime\prime}$ (6 pixels) for the 52 galaxies observed
at NOT and 2$^{\prime\prime} \times $2.3$^{\prime\prime}$ (3 pixels)
for the 8 galaxies observed at KPNO.  Using the distances in Table
\ref{tab:sample_properties}, the projected areas covered by these
extractions correspond to radii of 11--204 pc, with a median value of
71 pc considering only the LLAGN in the sample. The signal-to-noise
ratio was estimated from the rms fluctuation in the 4789--4839 \AA\
interval, which is free of major absorption or emission features. We
obtain $S/N$ in the 23--88 range, with a median value of 50. In the
4010--4060 \AA\ interval the median $S/N$ is 22.

\subsection{HFS extractions}

\label{sec:HFS_aperture}

Most of the HFS97 spectra were collected through a $2^{\prime\prime}
\times 4^{\prime\prime}$ aperture, $\sim 7$ times larger in area than
our typical nuclear extractions. In order to allow a more meaningful
comparison of our measurements with those in HFS97 we have also
extracted spectra through a $2^{\prime\prime} \times 4^{\prime\prime}$
aperture (hereafter the ``HFS-aperture''). This was straightforward
for the KPNO observations, which have a slit width equal to that in
HFS97.  In order to construct an HFS-aperture for the NOT data, which
was collected through a $1^{\prime\prime}$ slit, we first extracted a
$4^{\prime\prime}$ long spectrum and then multiplied it by 2 except in
the nucleus (ie, the inner $1.1^{\prime\prime}$), which was
represented by an interpolation between off-nuclear extractions
centered at $\pm 1.1^{\prime\prime}$ to avoid overweighting any
nuclear point source. The flux in this approximate HFS-aperture is
typically 4 times larger than in our nuclear extractions.  

All spectral properties discussed in \S\ref{sec:SpectralProperties}
were measured both through our nuclear extractions and through this
HFS-aperture, but only the nuclear values are published here, since
except for galaxies with noticeable gradients (discussed in Paper III)
the results are similar for both apertures. Tables for HFS-apertures
measurements are available upon request.

Besides spatial resolution, our observations differ from those of
HFS97 in spectral coverage. The blue spectra in HFS97 cover the
4230--5110 \AA\ interval, while our spectra go from 3500 up to 5500
\AA. As well shall soon see, the region between 3500 and 4200 \AA\
proved to be most revealing.

\section{Spectra}

In this section we present the spectra and two different methods to
characterize the stellar populations in the central regions of
LLAGN. The first method consists of an empirical, qualitative
classification based on the comparison with non-active galaxies
representative of simple stellar populations with different ages. The
second method quantifies this classification by modeling the starlight
using the non-active galaxies as spectral templates.  In Section 5 we
present measurements of several spectral indices (such as the 4000
\AA\ break, H$\delta$, CaII K, G band and others), which provide yet
another way to characterize stellar populations.  A comparison of all
these methods is important in view of the diversity of methods to
analyse the stellar populations of AGN found in the literature. This
comparison allows us to do a relative calibration between them, and
can be used to convert the results obtained with one method to the
others.

\label{sec:Spectra}

\subsection{Normal galaxies and an empirical stellar population classification}

\label{sec:spectra_Templates}

\begin{figure}
\caption{Spectra of the normal galaxies in the sample, including two
Starburst nuclei (top). Galaxies are sorted according to the age of
the dominant stellar population, with younger systems at the top. $Y$,
$I$, $I/O$ and $O$ denote our empirical stellar population classes.}
\label{fig:spectra_templates}
\end{figure}                                           

We start our presentation of the nuclear spectra by the 9 non-active
galaxies in our sample, shown in Figure \ref{fig:spectra_templates}. The
order of the spectra in this plot reflects approximately the age of
the dominant stellar population, younger systems at the top. These
normal galaxy spectra are used in this work as a reference basis for
an empirical stellar population classification. Galaxies in Figure
\ref{fig:spectra_templates} can be separated into four classes:

\begin{itemize}

\item[($Y$)] Galaxies with evidently young, $< 10^7$ yr, stellar
populations (NGC 3367 and NGC 6217), characterized by a blue continuum
and weak metal absorption lines;

\item[($I$)] galaxies with a dominant intermediate age, $10^8$--$10^9$ yr
population, characterized by pronounced HOBLs (NGC 205);

\item[($I/O$)] galaxies with a mixture of intermediate age and older
stars (NGC 221 and NGC 628);

\item[($O$)] galaxies dominated by an old stellar population (NGC 224,
NGC 1023, NGC 2950 and NGC 6654), characterized by strong metal lines.

\end{itemize}

For short, these four stellar population classes are denominated $Y$,
$I$, $I/O$ and $O$, respectively. We use the symbol $\eta$ to denote
these classes.

We have compared each of the LLAGN in our sample to these normal
galaxies and classified them onto these same 4 stellar population
categories according to a similarity criterion. This was first done by
a simple visual inspection of the spectra, and later confirmed by
means of the starlight modeling scheme described in
\S\ref{sec:StarlightModelling}. The results are listed in the last
column of Table \ref{tab:Template_Fitting}.  None of the objects
resembles predominantly young systems like NGC 3367, so all our LLAGN
fall onto the $\eta = I$, $I/O$ and $O$ categories.

This empirical classification provides a stellar population
counterpart to the LINER/TO classification, which is based on emission
line properties.  The $\eta$ classes are used here just as a heuristic
aid to sort the LLAGN spectra according to their main stellar
populations. This is a useful strategy to investigate connections
between emission line and stellar population properties. However,
classification criteria, while certainly helpful to identify general
trends, have a certain degree of uncertainty and arbitrariety. When
discussing our results we will thus not restrict ourselves to
comparisons between discrete emission line and stellar population
classes. In \S\ref{sec:Correlations} we give more emphasis to
comparisons between continuous stellar population and emission line
properties. We shall see that a clearer picture emerges if one avoids
the rigidity of taxonomy.

\subsection{LINERs and Transition Objects}

\label{sec:spectra_LINERs_ans_TOs}

The nuclear spectra for all LLAGN in our sample are shown in Figures
\ref{fig:spectra_LINERs_O}--\ref{fig:spectra_TOs_I}. Since our main
aim in this paper in to investigate the relation between the stellar
populations of these objects and their emission line properties, we
have split the spectra in six slots: strong-[OI] sources with either
$O$, $I/O$ and $I$ stellar population characteristics (Figures
\ref{fig:spectra_LINERs_O} and \ref{fig:spectra_LINERs_IO_and_I}) and
weak-[OI] with either $\eta = O$, $I/O$ and $I$ (Figures
\ref{fig:spectra_TOs_O_and_IO}, \ref{fig:spectra_TOs_IO_and_I} and
\ref{fig:spectra_TOs_I}).

As is typical of LLAGN, the spectra are completely dominated by
starlight. With few exceptions (eg, NGC 1052 and NGC 6501), emission
lines are generally weak, and sometimes altogether absent (eg, NGC 410
and NGC 5055). Important diagnostic lines like H$\beta$ and
[OIII]$\lambda\lambda$4959,5007 are rarely seen, although
[OII]$\lambda$3727 is detected in most spectra. As thoroughly
discussed and illustrated by HFS97, starlight subtraction is an
absolute necessity in order to measure emission lines in these
objects, so we postpone the analysis of emission lines to a future
communication. Except for [OII], lines fluxes for galaxies in the
present sample have already been measured by HFS97, including the
important region around H$\alpha$, so we will use their measurements
when necessary.  In the following we concentrate on the measurement
and analysis of stellar population features.

As is evident from Figures
\ref{fig:spectra_LINERs_O}--\ref{fig:spectra_TOs_I}, there is a clear
connection between stellar population characteristics and the emission
line properties encoded in the weak/strong-[OI] classification. Old
stellar populations are substantially more frequent in strong than in
weak-[OI] sources, whereas the opposite happens for stellar population
class $I$. Numerically, we find that out of our 17 strong-[OI] nuclei,
10 (59\%) are in the $\eta = O$ class, 5 (29\%) have $\eta = I/O$ and
only 2 (12\%) belong to the $\eta = I$ class, while the corresponding
numbers for the 34 weak-[OI]'s are 7 (21\%), 11 (32\%) and 16 (47\%)
respectively.

\begin{figure}
\caption{Nuclear spectra of strong-[OI] LLAGN with predominantly old
stellar populations ($\eta = O$).}
\label{fig:spectra_LINERs_O}
\end{figure}

\begin{figure}
\caption{Nuclear spectra of strong-[OI] LLAGN with stellar population
classes $\eta = I/O$ (a) and $\eta = I$ (b).}
\label{fig:spectra_LINERs_IO_and_I}
\end{figure}

\begin{figure}
\caption{Nuclear spectra of weak-[OI] LLAGN with stellar population
classes $\eta = O$ (a) and $\eta = I/O$ (b).}
\label{fig:spectra_TOs_O_and_IO}
\end{figure}

\begin{figure}
\caption{Nuclear spectra of weak-[OI] LLAGN with stellar population
class $\eta = I/O$ (a) and $\eta = I$ (b).}
\label{fig:spectra_TOs_IO_and_I}
\end{figure}

\begin{figure}
\caption{Nuclear spectra of weak-[OI] LLAGN with a strong intermediate
age stellar population ($\eta = I$).}
\label{fig:spectra_TOs_I}
\end{figure}

\subsection{High Order Balmer Absorption Lines}

\label{sec:HOBLs}

The most notable feature in this data set is the ubiquity of the HI
high order Balmer series in absorption (H8 and higher), particularly
among weak-[OI] nuclei. By definition, objects where these features
are clearly present are attributed $\eta = I$ (ie, they resemble NGC
205).  Systems with HOBLs are also bluer and have shallower metal
absorption lines than the rest. Overall, their spectra reveal clear
signs of a $10^8$--$10^9$ yr intermediate age, ``post-starburst''
population.  HOBLs may also be present in $I/O$ nuclei, but blended
with metal lines of old stars to such an extent that they do not stand
out clearly in the total spectrum.

HOBLs have been seen before in a few LLAGN, the most famous example
being NGC 4569 (Keel 1996), which is also in our sample (Figure
\ref{fig:spectra_TOs_I}a). However, an incidence rate as high as that
suggested by our survey had not been suspected before.  Ho \etal
(2003) argue that objects like NGC 4569 are not representative of
LLAGN in their sample, most of which have a ``demonstrably old''
stellar population. While we agree that NGC 4569 is an extreme example
and that old stars dominate the light in most LLAGN, our spectra
reveal the presence of intermediate age stars in at least 1/3 of the
objects. The reason why we identify more of these populations than Ho
\etal (2003) is that they had to base their judgment on the strength
of the H$\gamma$, H$\beta$ and H$\alpha$ absorptions, which are much
more affected by emission and dilution by old stars than HOBLs.

As remarkable as the high frequency of HOBLs in weak-[OI]'s is their
apparent dearth among strong-[OI] nuclei. HOBLs are $\sim 5$ times
more frequent in weak than in strong-[OI] sources ($\eta = I$
fractions of $\sim 50\%$ compared to $\sim 10\%$ respectively). These
findings reveal a strong link between the central stellar populations
and emission line properties, particularly [OI]/H$\alpha$.

The statistics above obviously depend on our adopted borderline
between weak and strong-[OI] nuclei (at [OI]/H$\alpha = 0.25$), but
the dichotomy persists (albeit somewhat diluted) even if one adopts
the HFS97 [OI]/H$\alpha = 0.17$ classification criterion.  A less
classification dependent assessment of the relation between stellar
and emission line properties is presented in \S\ref{sec:Correlations},
where we replace the discrete stellar population and emission line
classes used above by a more continuous description.

\subsection{Empirical starlight modeling}

\label{sec:StarlightModelling}

As noted in \S1, the location of TOs in between LINERs and HII-nuclei
in diagnostic diagrams suggests that their emission line spectrum
results from a mixture of AGN spectrum plus photoionization by young
stars (HFS97). If this is the case, one would expect to find direct
signatures of stellar populations of $10^7$ yr or less in their
spectra.  Although the high incidence rate of HOBLs among weak-[OI]
nuclei points to some sort of connection between stellar populations
and emission line properties, these features signal the presence of a
$\sim 10^8$--$10^9$ yr post-starburst-like component, not a young
starburst.  The clearest single signature of young stars in the
spectral range covered by our data is the WR bump at 4680 \AA.  This
feature is not evident in any of our LLAGN spectra. The WR bump was
only detected in the Starburst galaxy NGC 3367 (Figure
\ref{fig:spectra_templates}), which is part of our comparison sample.

Given the dominance of old and intermediate age stars in LLAGN, a weak
young starburst (with or without a correspondingly weak WR bump) could
well be lurking in some of our spectra. Detailed modeling of the
stellar population mixture is required to evaluate this possibility.
This modeling is performed in a separate communication (Paper II), in
order not to mix data with models. In this paper we want to keep the
analysis at a more empirical level. We have thus implemented an
entirely self contained starlight modeling technique which, instead of
resorting to theoretical spectra, uses only our observed spectra.

Each spectrum was modeled as a combination of a base of template
spectra. Two bases were considered: A ``normal galaxy base'',
containing five spectra from our comparison sample (NGC 3367, NGC 205,
NGC 221, NGC 1023 and NGC 2950), and an ``off-nuclear base'',
containing two off-nuclear extractions centered at $\pm
2.4^{\prime\prime}$ from the nucleus.  Except for the possibility of
using off-nuclear templates, this procedure is very similar to the one
employed by HFS97 in their starlight subtraction scheme.

All spectra were normalized to the flux at 4020 \AA\ and corrected for
Galactic extinction prior to the modeling.  The $A_B$ values of
Schlegel, Finkbeiner Davis (1998), extracted from NED\footnote{The
NASA/IPAC Extragalactic Database (NED) is operated by the Jet
Propulsion Laboratory, California Institute of Technology, under
contract with the National Aeronautics and Space Administration.}, and
the reddening law of Cardelli, Clayton \& Mathis (1989) were used for
this purpose. Intrinsic extinction by an uniform screen of dust was
allowed for. An algorithm was developed which searches for the
combination of base spectra and extinction which best reproduces the
nuclear spectrum of each LLAGN. The code was adapted from the
empirical population synthesis code of Cid Fernandes \etal (2001b) by
simply replacing its spectral base by the full spectra in either one
of these two bases. Regions around emission lines were masked out in
the comparison between observed and model spectra. For the off-nuclear
base, in most cases it was sufficient to mask out regions of 50 \AA\
around [OII], H$\beta$ and [OIII].  For the normal galaxy base we
further masked regions around H$\gamma$, H$\delta$ and
[NI]$\lambda$5200 since these are rather strong in the spectrum of NGC
3367, included in this base to represent a young starburst.

The presence of an emission line object in the normal galaxy base is
another difference with respect to the starlight modeling scheme
employed by HFS97. This difference is easily understood. HFS97 were
interested in measuring emission lines from a starlight free spectrum,
so their starlight templates were constructed out of galaxies without
emission lines. Our interest here is quite the opposite: We want to
analyze the starlight itself, not the emission lines. NGC 3367 was
introduced in the fitting to evaluate how strong a young starburst can
be accommodated in our LLAGN spectra. A consequence of this choice is
that our observed minus model residual spectra will have
underestimated, emission line fluxes when the NGC 3367 component is
present in a significant proportion. There is no way of circumventing
this without editing out the emission lines from NGC 3367 or resorting
to theoretical stellar population spectra, since an emission line free
starburst does not exist in nature. Nevertheless, in the majority of
cases we find that the NGC 3367 component contributes very little
($\le 2\%$) to the optical flux. In these cases, our residual spectra
can be seen as pure emission spectra.

\begin{figure}
\caption{Results of the starlight modeling for NGC 266 ($\eta = O$,
strong-[OI]), NGC 2681 ($\eta = I$, weak[OI]) and NGC 6500 ($\eta =
I/O$, weak-[OI]). {\it (a--c):} The top panels show the nuclear
spectrum $F_\lambda^{obs}$ (top) and two starlight models:
$F_\lambda^{\rm temp}$ (middle), constructed out of linear
combinations of the normal galaxies NGC 3367, NGC 205, NGC 221, NGC
2950 and NGC 1023; and $F_\lambda^{\rm off-nuc}$ (bottom), constructed
combining two off-nuclear extractions. All spectra are normalized to
the flux at 4020 \AA, and the two models are shifted downwards for
clarity. {\it (d--f):} $F_\lambda^{\rm obs} - F_\lambda^{\rm temp}$
residual spectra.  {\it (g--i):} $F_\lambda^{\rm obs} - F_\lambda^{\rm
off-nuc}$.}
\label{fig:starlight_subtraction}
\end{figure}

Examples of the results obtained with both spectral bases are shown in
Figure \ref{fig:starlight_subtraction}. The top panels show the
observed spectrum ($F_\lambda^{\rm obs}$, thick line) and the two
models for the starlight spectrum. The middle spectrum shows
$F_\lambda^{\rm temp}$, the template constructed out of linear
combinations of the normal galaxy base, while the bottom spectrum
($F_\lambda^{\rm off-nuc}$) is the combination of two off-nuclear
extractions which best matches the nuclear spectrum. Both models were
shifted vertically for clarity. The residual spectra are shown in the
bottom plots. Panels d--f correspond to $F_\lambda^{\rm obs} -
F_\lambda^{\rm temp}$, while panels g--i show $F_\lambda^{\rm obs} -
F_\lambda^{\rm off-nuc}$.  Except for regions containing emission
lines, the typical difference between $F_\lambda^{\rm obs}$ and the
$F_\lambda^{\rm temp}$ model is better than 3\% for all cases
shown. This is comparable to the noise level.  Off-nuclear templates
produce equally good results for the examples shown, with residuals of
4\% for NGC 266 and NGC 6500 and 1\% for NGC 2681. In some cases (eg,
NGC 772, NGC 4569, NGC 5921) the $F_\lambda^{\rm obs} - F_\lambda^{\rm
off-nuc}$ residual reveals a nuclear blue component not present in the
off-nuclear spectra. This happens specially in objects which have
strong young and/or intermediate age populations as revealed by the
strength of the NGC 3367 and NGC 205 components in the normal galaxy
decomposition.  We also note in passing that Figure
\ref{fig:starlight_subtraction} shows that HOBLs are present not only
in the nucleus of NGC 2681, but also in off-nuclear extractions. The
region producing the HOBLs is therefore {\it spatially extended}, a
result which is examined in more detail in Paper III.

One thus sees that, except for galaxies with noticeable gradients,
both methods yield comparable accuracy. For our current purposes, the
normal galaxy base is more relevant, since it provides a way of
quantifying the stellar population mix in LLAGN. In fact, this
decomposition method provides a quantitative basis for the comparative
stellar population classification outlined in
\S\ref{sec:spectra_Templates}.  To illustrate how this works, one can
write the relative contributions at $\lambda = 4020$ \AA\ of NGC 3367
($\eta = Y$), NGC 205 ($\eta = I$), NGC 221 ($\eta = I/O$) and NGC
2950 plus NGC 1023 ($\eta = O$) as a normalized four-component vector
${\bf x} = (x_Y,x_I,x_{I/O},x_O)$.  For the $\eta = O$ nuclei of NGC
266 and NGC 7626 we find ${\bf x} = (0,5,18,77)\%$ and $(1,2,0,97)\%$
respectively, while for NGC 4438, which we classified as $\eta = I/O$,
this vector is $(0,3,51,46)\%$, and for the $\eta = I$ nucleus of NGC
5921 we find $(2,70,4,24)\%$. In over 90\% of the cases this scheme
yields the same classification as that initially inferred by visual
inspection of the spectra. In the few cases where a disagreement
occurred, we adopted the visually estimated $\eta$ class, although
this has no consequence for the results of this paper.

The results of the starlight modeling in terms of the normal galaxy
base are summarized in Table \ref{tab:Template_Fitting}. A first
important result of this analysis is that 23 of our 51 LLAGN contain a
NGC 205 component stronger than 20\%. Except for NGC 5879, which has a
noisy spectrum, all nuclei where $x_I > 30\%$ were classified as $\eta
= I$ because of their conspicuous HOBLs. Between $x_I = 10\%$ and
$30\%$ some objects were classified as $\eta = I$ (NGC 772) and others
as $I/O$ (eg, NGC 4192, NGC 5055). The starlight analysis therefore
confirms the high incidence rate of HOBLs in LLAGN, specially
weak-[OI]'s, and helps quantifying the presence of these intermediate
age populations in cases where their contribution to the optical
spectrum is not visually obvious.

\begin{deluxetable}{lccccccc}
\tabletypesize{\scriptsize}
\tablewidth{0pc}
\tablecaption{Starlight modeling results}
\tablehead{
\colhead{Galaxy}     &
\colhead{$x_Y$}      &
\colhead{$x_I$}      &
\colhead{$x_{I/O}$}  &   
\colhead{$x_{O}$}    &
\colhead{$A_V$}      &
\colhead{$\Delta$}   &
\colhead{$\eta$}     \cr
\colhead{}           &
\colhead{(NGC 3367)} &
\colhead{(NGC 205)}  &
\colhead{(NGC 221)}  &
\colhead{(NGC 1023 + NGC 2950)} &
\colhead{}           &
\colhead{}           &
\colhead{}} 
\startdata
NGC 0266    &   0.1 &   5.2 &  17.8 &  77.0 & 0.18 &  2.8  &   O \cr
NGC 0315    &   3.8 &   9.2 &   1.5 &  85.5 & 0.09 &  3.5  &   O \cr
NGC 0404    &   0.0 &  78.5 &   0.0 &  21.5 & 0.90 &  6.6  & I   \cr
NGC 0410    &   0.0 &   1.8 &   0.7 &  97.5 & 0.04 &  2.6  &   O \cr
NGC 0428    &   0.0 &  49.8 &   0.0 &  50.2 & 0.31 &  8.7  & I   \cr
NGC 0521    &   1.3 &  11.1 &   0.1 &  87.6 & 0.31 &  2.7  &   O \cr
NGC 0660    &   0.0 &  59.4 &   0.0 &  40.6 & 2.38 & 11.7  & I   \cr
NGC 0718    &   0.0 &  51.3 &   2.3 &  46.4 & 0.13 &  2.7  & I   \cr
NGC 0772    &  32.1 &  10.9 &  24.4 &  32.5 & 0.38 &  1.8  & I   \cr
NGC 0841    &   0.0 &  37.7 &  17.4 &  44.9 & 0.18 &  2.3  & I   \cr
NGC 1052    &  16.5 &   0.0 &   0.5 &  83.0 & 0.57 &  3.0  &   O \cr
NGC 1161    &   0.0 &   6.7 &   0.1 &  93.2 & 0.13 &  2.7  &   O \cr
NGC 1169    &   0.0 &   0.0 &  48.2 &  51.8 & 0.26 &  5.7  & I/O \cr
NGC 2681    &   0.0 &  66.2 &   1.4 &  32.4 & 0.22 &  3.4  & I   \cr
NGC 2685    &   0.3 &   0.9 &  50.5 &  48.4 & 0.21 &  4.9  & I/O \cr
NGC 2911    &   0.0 &   3.4 &   0.0 &  96.6 & 0.83 &  4.6  &   O \cr
NGC 3166    &   0.0 &  28.8 &   9.4 &  61.8 & 0.19 &  2.3  & I/O \cr
NGC 3169    &   0.0 &  21.1 &   0.0 &  78.9 & 0.73 &  4.4  & I/O \cr
NGC 3226    &   0.0 &   2.7 &   0.0 &  97.3 & 1.08 &  4.2  &   O \cr
NGC 3245    &  14.4 &   3.8 &  25.5 &  56.3 & 0.17 &  1.5  & I/O \cr
NGC 3627    &   0.0 &  66.6 &   1.4 &  32.0 & 0.47 &  3.0  & I   \cr
NGC 3705    &   0.0 &  39.3 &  13.8 &  46.9 & 0.54 &  3.6  & I   \cr
NGC 4150    &   0.0 &  66.8 &  32.0 &   1.2 & 0.00 &  6.5  & I   \cr
NGC 4192    &   0.0 &  19.1 &  34.7 &  46.1 & 1.41 &  3.4  & I/O \cr
NGC 4438    &   0.0 &   2.6 &  50.9 &  46.5 & 1.48 &  2.9  & I/O \cr
NGC 4569    &  28.2 &  71.8 &   0.0 &   0.0 & 0.00 &  5.3  & I   \cr
NGC 4736    &   0.0 &  48.8 &   0.0 &  51.2 & 0.07 &  2.3  & I   \cr
NGC 4826    &   0.0 &  32.8 &  10.7 &  56.5 & 0.24 &  2.3  & I   \cr
NGC 5005    &   0.0 &  45.1 &   0.4 &  54.5 & 0.77 &  3.7  & I   \cr
NGC 5055    &   2.0 &  20.8 &   9.1 &  68.1 & 0.24 &  3.4  & I/O \cr
NGC 5377    &   0.0 &  76.8 &   0.1 &  23.1 & 0.02 &  2.5  & I   \cr
NGC 5678    &   0.0 &  89.7 &   0.3 &   9.9 & 1.22 &  3.8  & I   \cr
NGC 5879    &   0.0 &  52.0 &   0.9 &  47.1 & 0.46 &  7.3  & I/O \cr
NGC 5921    &   1.7 &  70.1 &   3.7 &  24.5 & 0.00 &  2.8  & I   \cr
NGC 5970    &   0.0 &   5.7 &  56.4 &  37.9 & 0.13 &  4.6  & I/O \cr
NGC 5982    &   0.2 &  10.3 &   2.5 &  87.0 & 0.02 &  1.9  &   O \cr
NGC 5985    &   0.0 &   6.8 &  34.4 &  58.8 & 0.19 &  4.9  &   O \cr
NGC 6340    &   0.0 &   4.1 &   0.0 &  95.9 & 0.98 &  6.6  &   O \cr
NGC 6384    &   1.2 &   2.2 &  61.3 &  35.3 & 0.02 &  4.4  & I/O \cr
NGC 6482    &  10.1 &   0.2 &   0.0 &  89.7 & 0.12 &  2.5  &   O \cr
NGC 6500    &  24.7 &   0.1 &   4.0 &  71.3 & 0.08 &  2.3  & I/O \cr
NGC 6501    &   1.2 &  11.5 &   0.0 &  87.3 & 0.25 &  2.3  &   O \cr
NGC 6503    &  22.5 &  55.8 &  16.3 &   5.4 & 0.32 &  3.0  & I   \cr
NGC 6702    &   0.0 &   7.0 &  18.6 &  74.4 & 0.00 &  2.7  &   O \cr
NGC 6703    &   0.0 &   3.4 &   0.0 &  96.6 & 0.22 &  3.5  &   O \cr
NGC 6951    &   2.0 &  29.4 &  18.9 &  49.7 & 0.41 &  3.1  & I/O \cr
NGC 7177    &   2.4 &  13.7 &  43.1 &  40.8 & 0.39 &  2.8  & I/O \cr
NGC 7217    &   0.0 &   7.2 &   0.0 &  92.8 & 0.43 &  3.6  &   O \cr
NGC 7331    &   0.0 &   5.3 &  15.0 &  79.8 & 0.37 &  2.8  &   O \cr
NGC 7626    &   1.1 &   1.8 &   0.0 &  97.2 & 0.09 &  2.2  &   O \cr
NGC 7742    &   0.0 &  14.3 &   0.0 &  85.7 & 0.16 &  4.1  & I/O \cr
\enddata
\label{tab:Template_Fitting}
\tablecomments{Results of the starlight decomposition in terms of the
normal galaxies NGC 3367, NGC 205, NGC 221, NGC 1023 and NGC 2950.
Cols.\ (2)--(5): $x_Y$, $x_I$, $x_{I/O}$ and $x_O$, given as
percentage fractions of the flux at 4020 \AA. Col.\ (6): V band
extinction (in mag) of the best fit. Col.\ (7): Mean absolute
percentage difference between the observed and model spectra: $\Delta
= <|F_\lambda^{\rm obs} - F_\lambda^{\rm model}| / F_\lambda^{\rm
obs}>$. Col.\ (8): Stellar population class.}
\end{deluxetable}

\subsection{The weakness of young stellar populations in LLAGN}

\label{sec:WRbump}

A second result of the starlight modeling is that few LLAGN have an
optically relevant young starburst. The contribution of the NGC 3367
component is larger than 20\% in just four cases. In order of
decreasing $x_Y$, these are NGC 772 (32\%), NGC 4569 (29\%), NGC 6500
(25\%) and NGC 6503 (22\%), which are all weak-[OI]'s. 

Of all our LLAGN, only NGC 6500 presents marginal evidence for the
presence of the WR bump. This can be seen by the coherent broad
residual with amplitude of no more than a few \% in the
$F_\lambda^{\rm obs} - F_\lambda^{\rm temp}$ residual spectrum in
Figure \ref{fig:starlight_subtraction}f. Naturally, the amplitude of
this residual increases a bit removing the WR bump from the NGC 3367
base component, but even then we cannot claim a conclusive detection
of this feature. One way to double check this possibility is to
evaluate the residual spectrum obtained with the off-nuclear base. If
WR stars are concentrated in the nucleus, a WR bump should appear in
the $F_\lambda^{\rm obs} - F_\lambda^{\rm off-nuc}$ difference
spectrum. In the case of NGC 6500, this residual reveals a weak narrow
nebular HeII$\lambda$4686, but no clear sign of a broad bump (Figure
\ref{fig:starlight_subtraction}i).  A tentative detection of broad
HeII in this galaxy has been previously reported by Barth \etal
(1997).

Detecting the WR bump is already hard in {\it bona fide} Starburst
galaxies (eg, Schaerer, Contini \& Kunth 1999), and the superposition
of a dominant older stellar population only makes matters worse.  In
NGC 3367 this feature has an equivalent width of $\sim 5$ \AA, similar
to the values in WR galaxies (Schaerer, Contini \& Pindao 1999) and
metal rich giant HII regions (Pindao \etal 2002). Hence, a NGC
3367-like component diluted to $< 30\%$ of the optical flux yields WR
bump equivalent widths of {\it at most} 1.5 \AA, which is comparable
to our spectral resolution, so weaker features are unlikely to be
detected.  The weakness of the young starburst component in LLAGN may
thus explain why we have not detected the WR bump in our data.

It is therefore clear that young stellar populations contribute very
little to the optical spectra of LLAGN.  It is important to point out
that even when $< 10^7$ yr stars dominate the UV light, as in NGC 404,
NGC 4569, NGC 5055 and NGC 6500 (Maoz \etal 1998), their contribution
does not exceed 30\% of the $\lambda4020$ flux. In fact, these young
starbursts are not even detected in the optical for NGC 404 and NGC
5055 due to their {\it low contrast} with respect to older
populations.

While current star formation seems to be proceeding at a residual
level in LLAGN, the ubiquity of intermediate age populations indicates
that its was much more prominent $10^8$--$10^9$ yr ago, which points
to a time-decaying star formation activity. This result contrasts with
that obtained for studies of type 2 Seyferts, where young stellar
populations are found in abundance (eg, Cid Fernandes \etal 2001a;
Joguet \etal 2001). The mean age of stars in the central regions of
Seyfert 2s is thus smaller than that in LLAGN. It is equally
instructive to compare the stellar populations of weak and strong-[OI]
LLAGN. The finding that intermediate age populations are found almost
exclusively among weak-[OI] nuclei implies that these sources are
on-average younger than those with strong [OI] emission. The
implications of these results for evolutionary scenarios are
considered in \S\ref{sec:Correlations} and Paper II.

\section{Spectral Properties}

\label{sec:SpectralProperties}

In this section we present measurements of a suite of stellar features
in the nuclear spectra presented above. Measurements through the HFS
aperture discussed in \S\ref{sec:HFS_aperture} are only used here for
comparison purposes. Tables with these and other measurements not
published here are available upon request.

\subsection{Equivalent widths and colors in Bica's system}

In a series of papers starting in the mid 80's, Bica \& co-workers
have explored a stellar population synthesis technique known as
Empirical Population Synthesis (EPS), which decomposes a given galaxy
spectrum in a combination of observed spectra of star clusters. In
practice, instead of modeling the $F_\lambda$ spectrum directly, this
method synthesizes a number of absorption line equivalent widths
($W_\lambda$) and continuum colors ($C_\lambda$), which are used as a
compact representation of $F_\lambda$. EPS has proven a very useful
tool in the analysis of galaxy spectra in a variety of contexts, from
normal galaxies (Bica 1988), to starbursts (Raimann \etal 2000; Cid
Fernandes, Le\~ao, Rodrigues-Lacerda 2003) and even AGN (Schmitt,
Storchi-Bergmann \& Cid Fernandes 1999; Raimann \etal 2003).

In this section we present measurements of $W_\lambda$ and $C_\lambda$
in Bica's system. These will be used here as a raw measures of stellar
population characteristics in our sample galaxies. These same data are
used in Paper II as input for an EPS analysis. As a side-step to this
investigation, we have developed a fully automated scheme to measure
$W_\lambda$ and $C_\lambda$ in this system. This is discussed next.

\subsubsection{Automated measure of the Pseudo Continuum}

A drawback of Bica's system is its subjectivity. Both equivalent
widths and colors are measured relative to a {\it pseudo continuum}
($PC_\lambda$), defined at pre-selected pivot wavelengths, which is
traced interactively over the observed spectrum (see Cid Fernandes
\etal 1998 for an illustrated discussion). Historically, this has
limited the use of EPS to a small number of researchers initiated in
the ``art'' of drawing this continuum.  Furthermore, the lack of an
objective recipe to define $PC_\lambda$ makes this whole system
useless in face of the huge spectral databases currently available,
which must be processed in an entirely automated manner. The present
data set is itself an example: Counting all nuclear and off-nuclear
extractions, there are over 700 spectra.

In order to remedy this situation, we have formulated an {\it a
posteriori} definition of Bica's system. This was done by means of a
straightforward computational procedure which mimics the placement of
the pseudo continuum, which until now has been carried out by hand.  

We started by measuring by hand a representative sub-set of 42 spectra
in the current data set, all normalized to the respective median flux
in the 4789--4839 \AA\ interval. For each pivot wavelength $\lambda$
we have then computed {\it median} fluxes in windows of different
sizes and locations. Windows were restricted to be at least 10 \AA\
wide and centered within 100 \AA\ of the corresponding pivot
$\lambda$. The manually measured $PC_\lambda$ fluxes were then
compared to the automatically measured median fluxes ($MF_\lambda$)
for all spectra, and a linear relation

\begin{equation}
\label{eq:AutoPC_definition}
PC_\lambda = a_\lambda + b_\lambda MF_\lambda 
\end{equation}

\noindent was fitted using an ordinary least squares. The optimal
window size and location where chosen to be those which produced the
best linear correlation coefficient (ie, smallest residuals).  Window
parameters and the corresponding $MF_\lambda$ to $PC_\lambda$
conversion coefficients $a_\lambda$ and $b_\lambda$ are listed in Table
\ref{tab:MF2PC_fits} for six pivot $\lambda$'s used in this work:
3660, 3780, 4020, 4510, 4630 and 5313 \AA.

\begin{deluxetable}{lrrr}
\tabletypesize{\scriptsize}
\tablewidth{0pc}
\tablecaption{Window definitions for the Pseudo Continuum}
\tablehead{
\colhead{Pivot $\lambda$ [\AA]} &
\colhead{Interval [\AA]}        &
\colhead{$a_\lambda^1$}           &
\colhead{$b_\lambda$}          }
\startdata
3660 & 3657--3667 &       0.0165  &  1.0069\cr
3780 & 3775--3785 &       0.0130  &  1.0274\cr
4020 & 4009--4020 &       0.0161  &  1.0019\cr
4510 & 4501--4512 &       0.0820  &  0.9337\cr
4630 & 4609--4627 &       0.1072  &  0.9037\cr
5313 & 5307--5317 &      -0.0437  &  1.0507\cr
\enddata
\label{tab:MF2PC_fits}
\tablenotetext{1}{The coefficient $a_\lambda$ is in units of the median
flux in the 4789--4839 \AA\ interval used to normalize the
spectra. $b_\lambda$ is adimensional}
\end{deluxetable}

This recipe works extremely well, with linear correlation coefficients
above $r = 0.97$ for all $\lambda$'s. Differences between manual and
automated $PC_\lambda$ measurements in our training set of 42 galaxies
are of order $3 \%$. This is well within the level of agreement
between manual measurements of $PC_\lambda$ carried out by different
people (Cid Fernandes \etal 1998). We have further verified that the
inclusion of other terms in equation (\ref{eq:AutoPC_definition}) does
not improve the fits significantly.  Figure \ref{fig:AutoPC_examples}
illustrates the pseudo continuum, as measured through our objective
system.  Equivalent widths are measured with respect to the continuum
formed by linear interpolation between the $PC_\lambda$ pivot
points. The differences between $W_\lambda$ measured with
manual and automatic $PC_\lambda$ is also of order 3\%, being always
smaller than 0.5 \AA.

We thus conclude that we have succeeded in formulating a non-subjective
definition of the pseudo continuum (and hence equivalent widths) in
Bica's system, thus overcoming a long held objection to this
measurement system.

\begin{figure}
\caption{Examples of the pseudo continuum in Bica's system, traced by
the automatic method developed here (see text). Solid dots mark the
pivot continuum points. Vertical dotted lines indicate windows of
integration of some of the absorption features in this system. From
left to right: $W_C$, $W_K$, $W_{CN}$, $W_G$ and $W_{Mg}$.}
\label{fig:AutoPC_examples}
\end{figure}                                           

\subsubsection{Results}

\label{sec:Results_Bica}

We have measured 7 equivalent widths in Bica's system: $W_C$,
$W_{wlb}$, $W_K$, $W_H$, $W_{CN}$, $W_G$ and $W_{Mg}$.  The window
definitions for these absorption features are the same defined in
previous works (Bica \& Alloin 1986; Bica, Alloin \& Schmidt 1994).
Note that $W_C$ is actually centered in a continuum region just to the
blue of H9. In old populations, the spectrum in this region is well
bellow the pseudo continuum due to a multitude of weak metal lines,
yielding $W_C \sim 4$--5 \AA\ (eg., NGC 266 and NGC 521 in Figure
\ref{fig:AutoPC_examples}). For populations of $\sim 1$ Gyr or less,
the continuum raises and $W_C$ approaches 0, as seen in NGC 4150 and
NGC 4569 (Figure \ref{fig:AutoPC_examples}). For our sample, this
index is a useful tracer of the presence or HOBLs. All cases where the
HOBLs are visually obvious have $W_C < 3$ \AA, while objects with $W_C
> 4$ \AA\ show no sign of these features.  $W_{wlb}$ (where ``wlb''
stands for ``weak line blend'') is centered on H9, but has a large
value (similar to the CaII K line, $W_K$) even in the absence of H9
because of the blend of many weak metal lines.  Detailed discussions
on the properties of these indices can be found in Bica \etal (1994)
and Storchi-Bergmann \etal (2000).

In Table \ref{tab:SpecIndices} we present the results for our nuclear
spectra. The continuum colors $PC_{3660} / PC_{4020}$ and $PC_{4510} /
PC_{4020}$, corrected for Galactic extinction, are also listed.  The
uncertainties in all these spectral indices were estimated by means of
Monte Carlo simulations. Each spectrum was perturbed 1000 times with
Gaussian noise with amplitude equal to the rms fluctuation in the
4010--4060 \AA\ interval. All indices were measured for each
realization of the noise, and the $1 \sigma$ uncertainties were
computed from the dispersion among the 1000 perturbed spectra. The
median uncertainties are between 0.3 and 0.5 \AA\ for all
$W_\lambda$'s and 0.03 for the $PC_{3660} / PC_{4020}$ and $PC_{4510}
/ PC_{4020}$ colors. The largest uncertainties are not more than twice
these values.

In Figure \ref{fig:Ws_X_RosaClass} we illustrate the relation between the
$\eta = Y$, $I$, $I/O$ and $O$ empirical stellar population classes
defined in \S\ref{sec:spectra_Templates} and the equivalents widths
$W_C$, $W_K$, $W_G$, $W_{Mg}$ and the $PC_{4510} / PC_{4020}$
color. The correspondence is excellent, as can be appreciated by the
locations of different symbols in all panels in this figure.  Nuclei
with young or intermediate age populations are confined to the low
$W_\lambda$ and blue color bottom-left corners of Figures
\ref{fig:Ws_X_RosaClass}a--d, while nuclei dominated by old stars
populate the high $W_\lambda$, red region, with $\eta = I/O$ objects
in between.  Approximate dividing lines can be placed at $W_C = 3.5$,
$W_K = 15$ $W_G = 9$, $W_{Mg} = 9$ \AA\ and $PC_{4510} / PC_{4020} =
1.55$.

Any of the $W_\lambda$'s in Figure \ref{fig:Ws_X_RosaClass} can therefore
be used as a continuous stellar population tracer which replaces our
discrete $I$, $I/O$, $O$ classification. In our previous study of the
stellar populations in type 2 Seyferts (Cid Fernandes \etal 2001a)
$W_K$ was used in this way. Of all absorption features measured here,
$W_K$ is the strongest one, which makes it also the more precise in
relative terms. Although the K line is actually produced by late type
stars, it indirectly traces younger populations by the dilution they
cause on $W_K$. For this reason, in practice $W_K$ is nearly as good a
tracer of HOBLs as a more direct measure of these features, like $W_C$
(Figure \ref{fig:Ws_X_RosaClass}a).

\begin{figure}
\caption{Relations between the equivalent widths of four absorption
lines and one color in Bica's system. Empirical stellar population
classes $Y$, $I$, $I/O$ and $O$ are represented by different symbols.
Squares, triangles, circles and stars represent $\eta = O$, $I/O$, $I$
and $Y$ respectively. (The two stars correspond to NGC 3367 and NGC
6217, two Starburst nuclei in our comparison sample.)  Dotted lines
approximately separate classes $\eta = Y$ and $I$ from $\eta = I/O$
and $O$. Mean error bars are indicated in the bottom right corner of
the figures.  }
\label{fig:Ws_X_RosaClass}
\end{figure}                                           

\begin{deluxetable}{lrrrrrrrrrrrr}
\tabletypesize{\tiny}
\tablewidth{0pc}
\tablecaption{Nuclear Spectral Properties}
\tablehead{
\colhead{Galaxy}        &
\colhead{$W_C$}         &
\colhead{$W_{wlb}$}     &
\colhead{$W_K$}         &
\colhead{$W_H$}         &
\colhead{$W_{CN}$}      &
\colhead{$W_G$}         &
\colhead{$W_{Mg}$}      &
\colhead{$\frac{F_{3660}}{F_{4020}}$} &
\colhead{$\frac{F_{4510}}{F_{4020}}$} &
\colhead{$D_n(4000)$}     & 
\colhead{H$\delta_A$}     &
\colhead{$W$(\ion{O}{2})} }
\startdata
NGC 0266 &   4.6 &  18.0 &  18.8 &  13.5 &  15.7 &  10.7 &  11.6 &  0.55 & 1.53 &    2.03 &   -2.1 &    5.4 \cr
NGC 0315 &   4.3 &  17.4 &  17.0 &  11.7 &  15.1 &  10.7 &  11.3 &  0.63 & 1.57 &    2.00 &   -2.8 &    8.1 \cr
NGC 0404 &   1.6 &   9.6 &   9.8 &  11.3 &   7.4 &   6.5 &   5.3 &  0.44 & 1.35 &    1.50 &    4.5 &    9.9 \cr
NGC 0410 &   4.9 &  19.4 &  17.6 &  12.9 &  17.6 &  10.6 &  12.9 &  0.60 & 1.55 &    2.13 &   -3.1 &   -1.0 \cr
NGC 0428 &   2.0 &  10.9 &  14.8 &  12.6 &   8.5 &   7.3 &   4.4 &  0.49 & 1.23 &    1.65 &    2.4 &   -1.7 \cr
NGC 0521 &   4.5 &  18.1 &  17.9 &  13.4 &  17.4 &  11.0 &  12.3 &  0.52 & 1.59 &    2.01 &   -2.7 &   -1.3 \cr
NGC 0660 &   1.3 &  10.8 &  14.5 &  11.5 &   7.3 &   7.5 &   7.1 &  0.47 & 1.71 &    1.72 &   -1.4 &   22.4 \cr
NGC 0718 &   2.6 &  13.1 &  13.1 &  12.9 &   9.5 &   8.3 &   7.7 &  0.52 & 1.30 &    1.65 &    3.1 &    1.0 \cr
NGC 0772 &   2.5 &  10.5 &  11.6 &  10.0 &  11.0 &   8.4 &   8.2 &  0.68 & 1.29 &    1.46 &   -0.3 &    0.8 \cr
NGC 0841 &   2.7 &  13.5 &  14.9 &  13.0 &  10.4 &   8.9 &   8.5 &  0.55 & 1.32 &    1.73 &    1.9 &    1.7 \cr
NGC 1052 &   4.5 &  15.8 &  17.3 &   8.2 &  18.1 &   9.9 &  11.1 &  0.63 & 1.61 &    1.88 &   -1.3 &   87.5 \cr
NGC 1161 &   5.0 &  19.4 &  19.0 &  13.8 &  18.5 &  11.6 &  12.2 &  0.56 & 1.63 &    2.16 &   -2.8 &   -1.7 \cr
NGC 1169 &   4.3 &  16.7 &  18.8 &  12.8 &  15.5 &  12.0 &  11.0 &  0.61 & 1.57 &    2.08 &   -2.4 &    7.3 \cr
NGC 2681 &   1.8 &  11.5 &  12.3 &  12.8 &   8.3 &   7.2 &   6.8 &  0.47 & 1.23 &    1.57 &    4.9 &    1.6 \cr
NGC 2685 &   5.4 &  17.6 &  18.7 &  13.6 &  15.0 &  11.6 &  10.1 &  0.55 & 1.53 &    2.01 &   -1.9 &    0.6 \cr
NGC 2911 &   4.2 &  18.4 &  17.7 &  13.5 &  17.7 &  10.9 &  11.2 &  0.51 & 1.68 &    2.05 &   -2.0 &   29.1 \cr
NGC 3166 &   3.2 &  15.2 &  15.9 &  13.0 &  12.1 &   9.4 &   8.5 &  0.53 & 1.39 &    1.80 &    0.8 &    4.7 \cr
NGC 3169 &   4.0 &  16.6 &  17.3 &  13.3 &  12.9 &   9.9 &   9.0 &  0.50 & 1.60 &    1.99 &   -0.4 &    6.2 \cr
NGC 3226 &   4.4 &  18.2 &  18.5 &  13.4 &  17.6 &  11.3 &  11.7 &  0.58 & 1.74 &    2.13 &   -2.4 &    7.7 \cr
NGC 3245 &   3.9 &  15.3 &  15.2 &  11.8 &  12.7 &   9.6 &  10.6 &  0.67 & 1.45 &    1.79 &   -1.5 &    1.7 \cr
NGC 3627 &   1.7 &  11.6 &  11.6 &  12.7 &   7.0 &   7.0 &   6.6 &  0.48 & 1.29 &    1.55 &    4.4 &    7.8 \cr
NGC 3705 &   3.0 &  12.9 &  14.8 &  13.1 &  10.1 &   8.7 &   7.6 &  0.49 & 1.40 &    1.73 &    2.2 &    4.0 \cr
NGC 4150 &   2.1 &  11.2 &  12.6 &  12.3 &   7.7 &   7.4 &   4.9 &  0.51 & 1.19 &    1.55 &    3.6 &    2.3 \cr
NGC 4192 &   3.0 &  15.3 &  16.0 &  10.9 &  11.8 &   9.0 &   9.5 &  0.53 & 1.76 &    1.93 &    0.1 &   13.4 \cr
NGC 4438 &   4.1 &  17.4 &  17.9 &  12.6 &  15.7 &  11.4 &   9.9 &  0.51 & 1.80 &    2.00 &   -1.9 &   57.8 \cr
NGC 4569 &   0.6 &   7.4 &   5.0 &   9.5 &   1.5 &   2.7 &   3.7 &  0.60 & 0.94 &    1.21 &    6.4 &    1.2 \cr
NGC 4736 &   2.6 &  13.6 &  12.9 &  12.3 &   9.9 &   8.4 &   7.6 &  0.53 & 1.29 &    1.67 &    2.8 &   -0.1 \cr
NGC 4826 &   2.7 &  15.2 &  14.4 &  12.7 &  13.2 &   9.6 &   9.1 &  0.52 & 1.41 &    1.76 &    1.0 &    4.6 \cr
NGC 5005 &   2.7 &  13.8 &  14.6 &  12.1 &  10.6 &   8.3 &   6.7 &  0.48 & 1.46 &    1.74 &    2.3 &   38.1 \cr
NGC 5055 &   3.8 &  15.1 &  14.1 &  11.9 &  15.1 &   8.7 &  10.5 &  0.54 & 1.47 &    1.75 &   -0.9 &   -1.3 \cr
NGC 5377 &   1.8 &  10.7 &   8.7 &  11.0 &   8.0 &   6.7 &   7.8 &  0.53 & 1.18 &    1.45 &    3.7 &    4.0 \cr
NGC 5678 &   1.1 &   7.8 &   8.7 &  10.3 &   3.8 &   6.0 &   5.6 &  0.49 & 1.40 &    1.48 &    3.6 &   -0.1 \cr
NGC 5879 &   3.2 &  11.7 &  16.3 &  12.7 &   8.8 &   8.7 &   7.5 &  0.48 & 1.33 &    1.72 &    0.8 &    3.2 \cr
NGC 5921 &   2.3 &  11.2 &  11.2 &  11.6 &   6.5 &   6.5 &   6.2 &  0.53 & 1.18 &    1.51 &    4.6 &    5.0 \cr
NGC 5970 &   4.1 &  14.5 &  18.4 &  13.3 &  10.1 &  10.2 &   8.0 &  0.61 & 1.40 &    1.86 &   -0.8 &   -2.5 \cr
NGC 5982 &   4.6 &  18.4 &  18.1 &  13.7 &  16.1 &  10.8 &  11.0 &  0.57 & 1.47 &    2.05 &   -1.9 &   -2.5 \cr
NGC 5985 &   4.0 &  15.5 &  18.9 &  12.1 &  13.2 &  10.6 &   9.6 &  0.59 & 1.51 &    1.97 &   -2.6 &   25.8 \cr
NGC 6340 &   4.9 &  18.1 &  20.0 &  14.3 &  17.1 &  12.1 &  11.2 &  0.49 & 1.92 &    2.26 &   -2.5 &    2.4 \cr
NGC 6384 &   4.9 &  15.2 &  18.6 &  12.9 &  10.3 &  10.7 &   9.9 &  0.66 & 1.45 &    1.92 &   -2.3 &   -2.7 \cr
NGC 6482 &   4.7 &  18.7 &  18.8 &  13.4 &  17.5 &  11.3 &  12.8 &  0.63 & 1.54 &    2.11 &   -2.7 &    4.4 \cr
NGC 6500 &   3.8 &  14.5 &  15.8 &   8.7 &  14.5 &  10.0 &  10.6 &  0.74 & 1.42 &    1.74 &   -4.3 &   98.2 \cr
NGC 6501 &   4.2 &  17.7 &  16.8 &  12.9 &  17.7 &  10.8 &  11.9 &  0.56 & 1.56 &    2.03 &   -2.2 &   -3.4 \cr
NGC 6503 &   1.3 &   7.9 &   9.5 &  10.4 &   4.3 &   5.8 &   6.2 &  0.67 & 1.15 &    1.36 &    2.9 &    4.5 \cr
NGC 6702 &   4.5 &  18.5 &  18.1 &  13.3 &  14.0 &  10.8 &  11.0 &  0.61 & 1.49 &    2.04 &   -1.7 &   -3.0 \cr
NGC 6703 &   4.8 &  18.7 &  18.5 &  13.8 &  16.3 &  11.5 &  11.5 &  0.56 & 1.60 &    2.11 &   -2.4 &   -1.9 \cr
NGC 6951 &   3.7 &  14.1 &  16.4 &  11.7 &  15.7 &  11.0 &  10.2 &  0.52 & 1.43 &    1.70 &   -0.2 &   21.3 \cr
NGC 7177 &   3.5 &  14.5 &  16.6 &  12.5 &  11.7 &  10.3 &   9.2 &  0.58 & 1.52 &    1.80 &   -1.3 &   11.2 \cr
NGC 7217 &   5.0 &  18.7 &  19.2 &  14.0 &  17.2 &  11.5 &  12.0 &  0.53 & 1.68 &    2.07 &   -2.6 &   12.2 \cr
NGC 7331 &   4.2 &  16.8 &  18.0 &  13.6 &  15.3 &  11.0 &  10.9 &  0.54 & 1.55 &    1.99 &   -1.0 &   -1.0 \cr
NGC 7626 &   4.9 &  19.2 &  18.1 &  13.9 &  17.9 &  11.0 &  12.6 &  0.58 & 1.58 &    2.12 &   -3.3 &   -1.2 \cr
NGC 7742 &   4.1 &  15.9 &  17.1 &  13.1 &  13.1 &  10.6 &   9.6 &  0.52 & 1.50 &    1.90 &   -0.9 &    2.2 \cr \hline
NGC 3367 &   0.3 &   1.8 &   2.6 &   1.1 &   1.5 &   1.4 &   1.6 &  1.07 & 0.86 &    1.05 &   -3.0 &   13.3 \cr
NGC 6217 &   0.3 &   3.9 &   2.5 &   5.3 &   2.7 &   2.2 &   2.3 &  0.75 & 0.88 &    1.13 &    3.1 &    5.9 \cr
NGC 0205 &   1.2 &   8.4 &   7.1 &  11.5 &   4.3 &   5.2 &   3.3 &  0.51 & 1.07 &    1.37 &    6.3 &   -2.0 \cr
NGC 0221 &   4.0 &  15.5 &  17.9 &  13.9 &  11.9 &  10.6 &   8.1 &  0.60 & 1.41 &    1.86 &   -0.9 &   -3.4 \cr
NGC 0224 &   4.8 &  18.6 &  17.5 &  13.7 &  18.5 &  10.3 &  12.0 &  0.63 & 1.54 &    2.06 &   -2.9 &   -4.1 \cr
NGC 0628 &   2.9 &  13.1 &  16.1 &  12.7 &   8.5 &   8.5 &   6.8 &  0.64 & 1.35 &    1.74 &    0.5 &   -2.4 \cr
NGC 1023 &   5.4 &  20.8 &  19.4 &  14.5 &  19.1 &  11.9 &  12.3 &  0.57 & 1.63 &    2.26 &   -3.7 &   -4.6 \cr
NGC 2950 &   4.6 &  18.7 &  17.4 &  13.2 &  17.0 &  10.6 &  11.3 &  0.59 & 1.48 &    2.04 &   -2.0 &   -3.0 \cr
NGC 6654 &   4.7 &  16.4 &  18.5 &  13.4 &  16.6 &  12.5 &  12.1 &  0.61 & 1.58 &    2.09 &   -1.6 &   -5.4 \cr
\enddata
\label{tab:SpecIndices}
\tablecomments{Col.\ (1): Galaxy name; Cols.\ (2--10): Equivalent
widths of seven absorption features and two colors, all in Bica's
system. Col.\ (11): 4000 \AA\ break index (Balogh \etal 1999).  Col.\
(12): H$\delta_A$ equivalent width of Worthey \& Ottaviani (1997).
Col.\ (13): [OII] equivalent width of Balogh \etal (1999). All
equivalent widths are given in \AA}
\end{deluxetable}

\subsection{Spectral Properties in HFS's system}

\label{sec:HFS_system}

HFS97 used a set of spectral indices (defined by Whorthey 1992 and
Held \& Mould 1994) which also allow a characterization of stellar
populations. Seven of the nine absorption equivalent widths and one of
the two colors measured by HFS97 are in the wavelength range covered
by our data: the G-band, H$\gamma$, Fe4383, Ca4455, Fe4531, Fe4668,
H$\beta$ and $c(44-49)$. We have measured these indices in our spectra
following the recipes in HFS97. This was done mainly for comparison
purposes.

In Figure \ref{fig:Compare_Ho_X_US}a we compare the equivalent width of
the G-band for our HFS-aperture spectra with the measurements of
HFS97. The mean difference is just 0.03 \AA, and the rms is 0.6 \AA,
within the measurement errors of both studies.  Similarly, Figure
\ref{fig:Compare_Ho_X_US}b compares the values of $c(44-49)$ in these
two studies. The difference in $c(44-49)$ is just $0.03 \pm 0.04$,
again within the errors.

In Figure \ref{fig:Compare_Ho_X_US}c we compare the equivalent widths
of the G band as measured in the HFS system and Bica's $W_G$, while in
Figure \ref{fig:Compare_Ho_X_US}d we plot the $c(44-49)$ color against
Bica's $W_K$. All quantities are measured from our nuclear spectra.
The relatively tight relations between these independent measurements
reinforces the impression from Figure \ref{fig:Ws_X_RosaClass} that
different stellar indices in galaxy spectra are highly correlated.  A
corollary of these correlations is that a single index (say, $W_K$) is
enough to provide a good first order characterization of the stellar
populations in our sample galaxies.  This relatively high degree of
redundancy in galaxy spectra is also detected by studies of principal
component analysis (e.g., Schmidt \etal 1991; Sodr\'e \& Cuevas 1997),
which show that a large fraction of the variance in galactic spectra
is described by a single principal component. The physical property
behind such an approximately mono-parametric behavior is the {\it age}
of the dominant stellar population (Schmidt \etal 1991; Ronen,
Aragon-Salamanca \& Lahav 1999).

The strong inter-relations between different stellar population
indices can be used to extend results drawn from our sample to the
full HFS sample. For instance, any correlation found between $W_C$ or
$W_K$ and emission line properties for our galaxies will also appear
with the $W$(G-band) HFS index, which has been measured by HFS97 for
most of their 162 LLAGN. Analogously, statistics related to our $Y$,
$I$, $I/O$ and $O$ stellar population classes can be readily extended
to the full HFS sample exploring the relation between $\eta$ and, say,
$W$(G-band).

\begin{figure}
\caption{{\it Top:} (a) Comparison of the equivalent width of the G
band and (b) the $c(44-49)$ color for our HFS-aperture spectra and the
measurements of HFS97. Both indices were computed according to the
definitions in HFS97. Diagonal lines indicate $y=x$ and the $\pm 2$
sigma range. {\it Bottom:} Relation between $W_G$ and $W_K$, measured
through Bica's system, and two HFS indices: (c) the G-band equivalent
width and (d) $c(44-49)$. Symbols as in Figure \ref{fig:Ws_X_RosaClass}.}
\label{fig:Compare_Ho_X_US}
\end{figure}                                           

\subsection{Other spectral indices: $D_n(4000)$, H$\delta_A$ and 
$W$(\ion{O}{2})}

\label{sec:D4000}

Besides the CaII K line and the HOBLs, the interval between 3500 and
4200 \AA\ contains other features of interest.  The 4000 \AA\ break
and H$\delta$, for instance, have been recently used as stellar
population diagnostics for SDSS spectra of both normal and active
galaxies (Kauffman \etal 2002, 2003).  We have measured these two
indices for our spectra following the Balogh \etal (1999) definition
of $D_n(4000)$ and the Worthey \& Ottaviani (1997) definition of
H$\delta_A$. These are the same indices used in the SDSS work. We have
not corrected H$\delta_A$ for emission as this is insignificant for
most of our LLAGN. We further measured the [OII] equivalent width
$W$(\ion{O}{2}) according to the recipe of Balogh \etal (1999). All
spectra were corrected for Galactic reddening prior to these
measurements.

These indices are listed in the last three columns of Table
\ref{tab:SpecIndices}. Note that H$\delta_A > 0$ corresponds to
absorption while $W$(\ion{O}{2}) $> 0$ corresponds to emission. Note
also that negative $W$(\ion{O}{2}) is obtained in many cases, but this
does not necessarily means a non-detection of [OII]. For instance,
[OII] is clearly seen in the starlight subtracted spectrum of NGC
7626, while the $W$(\ion{O}{2}) index, which is measured over the
total spectrum, yields $-1.2$ \AA.  As for other emission lines in
LLAGN, the accurate measurement of [OII] requires a careful starlight
subtraction.

In Figure \ref{fig:D4000_Hd_OII}a we examine the relation between
$D_n(4000)$ and H$\delta_A$. The global distribution of LLAGN in this
diagram is similar to that obtained for SDSS galaxies (Kaufmman \etal
2002). Interestingly, our $\eta = I$ LLAGN, which are nearly all
weak-[OI] sources, lie above the locus of steady star formation models
of Kaufmann \etal (2002), in a region occupied by galaxies which have
experienced bursts of star formation in the past $\sim 10^9$ yr. This
is consistent with the strong HOBLs found in these systems and the
considerations of \S\ref{sec:WRbump}.  The location of $\eta = I$
weak-[OI] sources in Figure \ref{fig:D4000_Hd_OII}a is also similar to
that of AGN with L$_{\rm [OIII]} < 10^7$ L$_\odot$ and TO-like
emission line ratios in Kauffman \etal (2003). Notwithstanding these
similarities, it is important to observe that while the SDSS work
pertains to scales of several kpc, our data traces stellar populations
on scales of typically 70 pc.

Figure \ref{fig:D4000_Hd_OII}b illustrates the strong relation between
Bica's $W_K$ and $D_n(4000)$, which is not surprising given that the
$W_K$ window is contained in blue window of $D_n(4000)$.  The dashed
line indicates the quadratic fit:

\begin{equation}
\label{eq:Dn_x_WK_fit}
D_n(4000) = (1.013\pm0.006) + 
            (0.031\pm0.001) W_K + 
            (0.00124\pm0.00007) W_K^2
\end{equation}

\begin{figure}
\caption{Relations between $D_n(4000)$, H$\delta_A$ and $W_K$. Symbols
as in Figure \ref{fig:Ws_X_RosaClass}. The dashed line in (b) shows a
second order polynomial fit.}
\label{fig:D4000_Hd_OII}
\end{figure}                                           



\section{The Link Between Stellar Population and Emission Line 
Properties in LLAGN}

\label{sec:Correlations}

Several times in \S\ref{sec:Spectra} we have noted an apparent
connection between stellar population features revealed by our
spectroscopy and the weak/strong [OI] classification. In this section
we use the spectral properties measured in
\S\ref{sec:SpectralProperties} in conjunction with the [OI]/H$\alpha$
ratio from HFS97 to investigate this connection in a quantitative
manner.

One way to investigate this relation more closely is to correlate
[OI]/H$\alpha$ with the $W_C$ index, a useful tracer of the presence
of HOBLs. The strong tendency of HOBLs to appear preferentially in
weak-[OI] nuclei noted in \S\ref{sec:HOBLs} should reveal itself as a
correlation between [OI]/H$\alpha$ and $W_C$. Similar correlations
should exist between [OI]/H$\alpha$ and the equivalent widths of metal
absorption lines, since HOBLs come with an associated blue continuum
which dilutes spectral features like the K line and the G band.

Figure \ref{fig:Ws_x_O1Ha} confirms these expectations. The figure plots
[OI]/H$\alpha$ against our nuclear measurements of $W_C$, $W_K$, $W_G$
and $W_{Mg}$ (Table \ref{tab:SpecIndices}). What is seen in these
plots is not so much a correlation in the sense of points scattered
around a line, but a marked {\it dichotomy}: Nearly all nuclei with
weak absorption lines are weak-[OI] emitters, whereas nuclei with
strong metal lines span the full range of [OI]/H$\alpha$.  In
graphical terms, the top-left regions of all panels in Figure
\ref{fig:Ws_x_O1Ha} are remarkably empty!

\begin{figure}
\caption{Comparison of absorption equivalent widths, which trace
stellar populations, with the [OI]/H$\alpha$ ratio, an AGN indicator.
The [OI]/H$\alpha = 0.25$ border-line which separates weak from
strong-[OI] nuclei is indicated. The vertical dotted lines are the
same as in Figure \ref{fig:Ws_X_RosaClass}; they separate objects with a
strong intermediate age population ($\eta = I$) from objects dominated
by old stars ($\eta = O$). Symbols as in Figure
\ref{fig:Ws_X_RosaClass}.}
\label{fig:Ws_x_O1Ha}
\end{figure}                                           

This is not the first time in this paper that [OI]/H$\alpha$ is
plotted against a stellar population index. In Figure
\ref{fig:OurSample_X_HFSSample_O1Ha_WG} we plotted [OI]/H$\alpha$
versus $W$(G band) for the full HFS sample. There we see the same
``inverted L'' shape of any of the panels of Figure
\ref{fig:Ws_x_O1Ha}. Overall, these figures spell out a clear message:
The Narrow Line Region of LLAGN evidently knows what value of
[OI]/H$\alpha$ it can or cannot assume for a given stellar
$W_\lambda$. This finding implies a strong connection between stellar
populations and gas excitation processes in LLAGN.

As suggested by the referee, an alternative reading of this result
could be that galaxies with important circumnuclear star formation
have an excess of H$\alpha$ emission in the central few arcsec (note
that the [OI]/H$\alpha$ values are from the $2^{\prime\prime} \times
4^{\prime\prime}$ measurements by HFS97), resulting in a lower
[OI]/H$\alpha$ ratio.  We have studied the aperture dependence of the
H$\alpha$ flux in order to check whether the link between
[OI]/H$\alpha$ and the stellar population is not a consequence of an
excess of H$\alpha$ flux in the larger HFS97 apertures, compared to
the smaller apertures ($1^{\prime\prime} \times 1.1^{\prime\prime}$)
used in this work to derive the stellar population properties. We have
retrieved from the HST archive STIS (G750M or G750L) spectra of 27 out
of the 28 LLAGN analysed in Paper II.  We find that H$\alpha$ is very
compact in all strong-[OI] LLAGN.  In weak-[OI] LLAGN, the geometry of
H$\alpha$ is more diverse: in 7/22 H$\alpha$ is compact, and in 14/22
more than 50$\%$ of the flux (measured through a $0.2^{\prime\prime}
\times 4^{\prime\prime}$ extraction) is inside the central
$0.2^{\prime\prime} \times 1^{\prime\prime}$.  On the other hand, we
find that the extension of the H$\alpha$ emission does not correlate
with the type of nuclear stellar population.  In addition, we find
that there is no correlation between the H$\alpha$ luminosity and the
[OI]/H$\alpha$ ratio (both measured by HFS97).  We thus conclude that
the lower values of [OI]/H$\alpha$ in weak-[OI] LLAGNs are not due to
an excess of H$\alpha$ emission provided by recent star formation in
spatial scales of a few arcsecs.


A compelling visualization of the relation between [OI]/H$\alpha$ and
the stellar population properties is presented in Figure
\ref{fig:O1Ha_x_W_histograms}. In this figure we divide objects into
bins in stellar $W_\lambda$ and within each bin we compute the
fraction of objects with [OI]/H$\alpha < 0.17$ (filled areas), $0.17
\le$ [OI]/H$\alpha \le 0.25$ (hatched areas) and [OI]/H$\alpha > 0.25$
(empty areas). The panels with $W_K$ and $W_C$ are for our sample,
while the panel with $W$(G-band) uses data tabulated in HFS97 for all
their LLAGN.

The predominance of weak [OI] emitters among objects with low stellar
absorption equivalent widths is obvious to the eye. Of the 19 LLAGN
with $W_K < 15$ \AA, 17 have [OI]/H$\alpha \le 0.25$. Similarly 18 of
the 21 nuclei with $W_C < 3.5$ \AA\ have [OI]/H$\alpha \le
0.25$. Hence, $\sim 90\%$ of objects with weak absorption lines are
also weak [OI] emitters. Note that the above limits in $W_K$ and $W_C$
correspond approximately to the border line between our $\eta = I$ and
$\eta = I/O$ or $O$ stellar population classes (Figure
\ref{fig:Ws_X_RosaClass}), so nearly identical statistics are obtained
comparing $\eta$ with [OI]/H$\alpha$. Likewise, since $\eta = I$
systems are also those with HOBLs, one can alternatively express the
above statistics in terms of a [OI]/H$\alpha$-HOBLs connection.

For the full HFS sample, $\sim 80\%$ of the $W$(G band) $< 4.5 $ \AA\
sources have [OI]/H$\alpha \le 0.25$ (Figure
\ref{fig:O1Ha_x_W_histograms}c). This same fraction results if one
uses the HFS97 measurements for the 51 LLAGN in our sample, which
reinforces our conclusion that there are no significant differential
selection effects between the two samples.

It is important to draw attention to the fact that although most
objects with HOBLs and diluted metal lines are weak in [OI], {\it the
converse is not true}.  There is a numerous population of weak-[OI]
sources {\it without} conspicuous HOBLs and with stellar population
indices typical of old, $\sim 10$ Gyr populations. Among objects with
$W_K > 15$ \AA, for instance, $\sim 50\%$ have [OI]/H$\alpha \le
0.25$.  These are the objects located at the bottom-right corners of
the ``inverted L'' distributions in Figures
\ref{fig:OurSample_X_HFSSample_O1Ha_WG} and \ref{fig:Ws_x_O1Ha}, also
represented by the filled and hatched areas in the large $W_\lambda$
bins in Figure \ref{fig:O1Ha_x_W_histograms}.

\begin{figure}
\caption{Fraction of objects in each of three ranges in [OI]/H$\alpha$
as a function of different stellar absorption equivalent widths.
Filled areas are used for [OI]/H$\alpha < 0.17$, hatched areas for
$0.17 \le$ [OI]/H$\alpha \le 0.25$ and empty boxes for [OI]/H$\alpha >
0.25$. Numbers at the top indicate the number of objects in each
$W_\lambda$-bin. The left and middle panels are for the present sample
of 51 LLAGN, while the right panel is for the full HFS sample.}
\label{fig:O1Ha_x_W_histograms}
\end{figure}                                           

\subsection{Interpretation}

\label{sec:interpretation}

The shape of the distribution of points in Figure \ref{fig:Ws_x_O1Ha}b
lends itself to division into four regions, delimited by [OI]/H$\alpha
\sim 0.25$ in the vertical (emission line) axis and $W_K \sim 15$ \AA\
in the horizontal (stellar population) axis.  Similar divisions can be
placed at $W_C \sim 3.5$, $W_G \sim 9$ and $W_{Mg} \sim 9$ \AA\ in the
other panels of Figure \ref{fig:Ws_x_O1Ha}, and at $W$(G-band) $\sim 4.5$
\AA\ in Figure \ref{fig:OurSample_X_HFSSample_O1Ha_WG}. The top-left of
these four quadrants is essentially empty. What kind of objects
populate the remaining three quadrants?

Objects with clear signs of stellar populations of $\sim 10^9$ yr or
less (ie, those with HOBLs, diluted metal lines and blue colors) all
live in the bottom-left quadrant. They are all weak-[OI]'s according
to our definition. A plausible reading of this fact is that in these
objects stars are directly linked to the gas excitation by some as yet
undefined mechanism (see \S\ref{sec:Introduction} for a menu of
possibilities), resulting in higher H$\alpha$ per [OI] photon ratio
than in strong-[OI] LINERs. These sources could be called {\it
``Young-TOs''}, where ``young'' actually refers to populations of
$10^9$ yr of less. Note that the stellar $W_\lambda$'s in these
sources are intermediate between those of starburst and early type
galaxies. In other words, they are transition objects also in terms of
their stellar populations.

In the opposite end of the [OI]/H$\alpha$ versus $W_\lambda$ diagrams,
stars cannot play an important role in the gas excitation, since
photoionization by hot stars, be they young (ie, a starburst) or old
(post-AGB stars), produces smaller values of [OI]/H$\alpha$ than
observed in AGN.  The absence of optically significant $\la 10^9$ yr
populations corroborates this interpretation. Photoionization by an
AGN is likely the dominant emission line mechanism in these
sources. Variations in nebular conditions or in the ionizing spectrum
can be invoked to explain the range of [OI]/H$\alpha$ values spanned
by the data. A fitting denomination for nuclei in this top-right
quadrant is {\it ``Old-LINERs''}.

The bottom right quadrant of Figure \ref{fig:Ws_x_O1Ha} is populated
by objects with TO-like emission line spectra but no strong optical
signs of young or intermediate age stellar populations. These {\it
``Old-TOs''} may well belong to the Old-LINER family, in which case
their weak [OI]/H$\alpha$ should perhaps be explained by an ionizing
radiation field softer than for strong-[OI] LINERs.  Alternatively,
they may be simply weak Young-TOs whose $\la 10^9$ yr stars are
completely overwhelmed by the older population.  This is likely the
case of NGC 5055. The young starburst which clearly dominates the UV
light of this galactic nucleus (Maoz \etal 1998) leaves no trace in
the optical (Table \ref{tab:Template_Fitting}), although our starlight
decomposition indicates that 21\% of the 4020 \AA\ flux originates in
an intermediate age population. This non-negligible contribution,
however, causes little dilution of the metal lines. The lower right
quadrant of Figure \ref{fig:Ws_x_O1Ha} probably contains other such
objects.

To summarize, our results suggest a scenario where stars younger than
$\sim 1$ Gyr somehow participate in the gas ionization of Young-TOs,
whereas in Old-LINERs an AGN is the dominant ionizing agent. The
situation is less clear for Old-TOs, which may comprise a mixture of
the above objects. Interestingly, ``Young-LINERs'' do not seem to
exist. AGN taxonomy surely does not need more sub-classes. Yet, LLAGN
have long been known to comprise a heterogeneous class, and, as
discussed in \S\ref{sec:Introduction}, models involving AGN- and
stellar-driven ionization can both explain their emission line
properties. This empirically motivated scenario comes in the sense of
trying to elucidate which of these two broad physical processes, AGN
or stars, best fits a given source.  It is equally important to
clarify that we do not claim Young-TOs to be AGN-less. After-all, the
current common wisdom is that all galaxies with bulges harbor an AGN,
be it dead, alive or somewhere in between. In Young-TOs the AGN
contribution to the gas ionization would be small. Conversely,
ionization by stellar processes in Old LINERs would be overwhelmed by
the AGN.

A potential caveat in identifying nuclei with weak metallic lines with
stellar processes is that dilution by a quasar-like non-stellar
continuum can also produce the same effect. Presumably, this should be
more important for type 1 LLAGN, ie, L1.9 and T1.9 in the notation of
HFS97. This effect indeed happens in HST spectra of some objects
(Paper II). Interestingly, the two objects with $W$(G band) between
3.5 and 4 \AA\ and [OI]/H$\alpha \sim 0.6$ in Figure
\ref{fig:OurSample_X_HFSSample_O1Ha_WG} which intrude slightly into
the ``zone of avoidance'' are NGC 841 and NGC 5005, both classified as
L1.9.  Although it is conceivable that this effect operates in a few
cases, we find that type 1 and type 2 sources are well mixed in all
our diagrams, so this cannot be a major effect in our ground based
observations. Furthermore, we point out that HOBLs were detected in
both NGC 841 and NGC 5005, as well as in NGC 2681, another L1.9 with
relatively weak metal absorption lines.  It therefore seems more
likely that the main diluting agent even in these type 1 LLAGN is the
continuum associated with stellar populations of $\la 10^9$ yr rather
than with a quasar-like continuum.

\subsection{Evolution and Contrast}

As time passes, HOBLs and the continuum of $\la 10^9$ yr populations
gradually disappears, and metal absorption features become
progressively stronger. Passive evolution thus moves nuclei in left to
right direction in Figures \ref{fig:OurSample_X_HFSSample_O1Ha_WG} and
\ref{fig:Ws_x_O1Ha} on time scales of a few Gyr.  In the notation of
the previous section, objects in the left half of Figures
\ref{fig:OurSample_X_HFSSample_O1Ha_WG} and \ref{fig:Ws_x_O1Ha} are
nearly all Young TOs. If [OI]/H$\alpha$ remains constant during this
period, stellar evolution would gradually turn a Young-TO onto an
Old-TO. This would correspond to a horizontal evolutionary track onto
any of our [OI]/H$\alpha$ versus $W_\lambda$ diagrams.

However, the hypothesis of constant [OI]/H$\alpha$ does not seem
warranted by the ``inverted L'' distribution of LLAGN in Figures
\ref{fig:OurSample_X_HFSSample_O1Ha_WG} and \ref{fig:Ws_x_O1Ha}, which
suggests that the presence of $\la 10^9$ yr populations affects the
[OI]/H$\alpha$ ratio. For didactic purposes, lets consider a situation
in which an AGN is entirely responsible for the production of [OI]
photons, while H$\alpha$ is produced both by the AGN and $\la 10^9$ yr
stars.  In this case the stellar contribution to [OI]/H$\alpha$ acts
just in the denominator of this ratio. Evolution would thus move
objects both to the right and upwards along curved mixing lines in
Figures \ref{fig:OurSample_X_HFSSample_O1Ha_WG} and \ref{fig:Ws_x_O1Ha}
as the stellar powered H$\alpha$ emission fades. Considering that the
AGN part of [OI]/H$\alpha$ can assume a range of values (due, eg, to
different nebular conditions), one arrives at a scenario where stellar
evolution takes points in the bottom-left to any possible height in
the right part of these figures. In other words, Young-TOs can end up
as either Old-TOs or Old-LINERs.

It thus seem plausible that evolution plays a role in defining the
looks of LLAGN spectra, both in their stellar and emission lines
properties. Physical models for the evolution of [OI]/H$\alpha$ must
bear in mind that the intensity of [OI] is not only driven by the
ionizing radiation field; it also depends strongly of the nebular
geometry. Thus, variations of the ionization parameter and/or the
density structure would also affect the [OI]/H$\alpha$ ratio. Such
variations may be tied to the evolution of stellar populations. In
starbursts, for instance, [OI]/H$\alpha$ is expected to increase with
age as the bubble produced by the action of stellar winds and SNe
expands (Stasin\'ska, Schaerer \& Leitherer 2001; Stasin\'ska \&
Izotov 2003).

Besides evolution, contrast effects must also be present, as can be
illustrated by the case of NGC 6500. Judging by the relative
contributions of the $Y$, $I$ and $I/O$ components in our starlight
decomposition (25, 0 and 4\% of the 4020 \AA\ flux, respectively),
this is the youngest LLAGN in our sample. Yet, since 71\% of the
optical light comes from old stars, this galaxy does not quite make it
into our observational definition of Young-TO, whereas from the above
discussion of evolutionary effects one could naively expect it to be
located at the bottom left of our [OI]/H$\alpha$ versus $W_\lambda$
diagrams. Further examples of contrast effects are NGC 5055 and NGC
404, whose optical spectra reveal no sign of the young massive stars
unambiguously detected at UV wavelengths by Maoz \etal (1998). This
starburst component is simply buried under the much stronger
contribution of intermediate age and older stars.

A quantitative assessment of these ideas is beyond the scope of this
paper. Future communications will explore these issues by means of
stellar population synthesis, spectral energy distribution modeling
and comparisons to Seyfert and Starburst galaxies.

\section{Conclusions}

\label{sec:Conclusions}

In this paper we presented the first results of a survey of the
stellar populations of low luminosity AGN. Nuclear spectra covering
the inner $\sim 70$ pc in radius were presented for 51 LINERs and
LINER/HII Transition Objects plus a comparison sample 9 non-active
galaxies. The stellar population properties of LLAGN were quantified
by means of a comparison with non-active galaxies as well as by
measurement of a suite of spectral indices over the 3500--5500 \AA\
interval.

Our main results can be summarized as follows:

\begin{enumerate}

\item Young, $< 10^7$ yr starbursts contribute very little to the
optical spectrum of LLAGN. Even in the few nuclei known from UV
observations to contain young massive stars, these populations
participate with no more than 30\% of the optical continuum light, and
sometimes are altogether undetected due to a much stronger
contribution from other stellar populations.

\item Intermediate age, $10^8$--$10^9$ yr populations, on the other
hand, are found in at least 1/3 of all LLAGN. These populations are
easily characterized by pronounced HI Balmer lines in absorption,
diluted metal lines and relatively blue colors.

\item We found a very strong empirical connection between the stellar
populations and emission line properties of LLAGN.  Nearly all nuclei
containing substantial populations of young and/or intermediate age
stars are weak [OI] emitters, with [OI]/H$\alpha \le 0.25$, whereas
only $\sim 10\%$ of the sources with stronger [OI] contain such
populations. The mean stellar age of weak-[OI] LLAGN is thus several
Gyr younger than that of strong-[OI] LLAGN.  Nuclei dominated by older
stars, however, span the full range of [OI]/H$\alpha$ values, from TOs
to LINERs.

\item Given a few Gyr, TOs with conspicuous intermediate age
populations will naturally evolve to either ``Old-TOs'' or
``Old-LINERs''.

\end{enumerate}

Perhaps the major lesson from this study is that the information
contained in the stellar populations of active galaxies can provide
valuable clues on the nature of these systems. Combining this
information with emission line data, we were able to establish an
inequivocal connection between stellar populations and the gas
excitation mechanism in LLAGN. The physical origin of this connection,
however, has not been unveiled in this work. By analogy with previous
investigations (particularly of type 2 Seyferts), starburst activity
is the prime contender to explain the relation between [OI]/H$\alpha$
and stellar population properties reported in this work. Direct
evidence for young stars in LLAGN is nevertheless too scant to fully
subscribe this interpretation at this stage, specially because other
scenarios (such as SN shocks and photoionization by evolved
post-starburst populations) may potentially explain our
findings. Future papers in this series will address these and other
related issues exploring complementary spectroscopic and imaging data,
as well as detailed modeling of the stellar populations in these
sources.

\acknowledgements It is a pleasure to thank Grazyna Stasin\'ska, Luis
Colina and \'Aurea Garcia for their enlightening comments and
suggestions, and the referee for her/his suggestions that help to improve the paper.
  RCF and TSB acknowledge the support from CNPq and
CAPES. RGD and EP acknowledge support by Spanish Ministry of Science
and Technology (MCYT) through grant AYA-2001-3939-C03-01.  The data
presented here have been taken using ALFOSC, which is owned by the
Instituto de Astrof\'{\i}sica de Andaluc\'{\i}a (IAA) and operated at
the Nordic Optical Telescope under agreement between IAA and the
NBIfAFG of the Astronomical Observatory of Copenhagen. We are very
grateful to the IAA director for the allocation of 5.5 nights of the
ALFOSC guaranteed time. The National Radio Astronomy Observatory is a
facility of the National Science Foundation operated under cooperative
agreement by Associated Universities, Inc.



\end{document}